\RequirePackage{fix-cm}
\documentclass[smallcondensed]{svjour3}     % onecolumn (ditto)
\smartqed  % flush right qed marks, e.g. at end of proof
\usepackage{graphicx}
\usepackage{graphicx,epstopdf}% Include figure files
\usepackage{dcolumn}% Align table columns on decimal point
\usepackage{bm}% bold math

\begin{document}           % End of preamble and beginning of text.

\title{Cosmological General Relativity With Scale Factor and Dark Energy}

\author{Firmin J. Oliveira \\
%P. O. Box 10882\\ Hilo, Hawai`i, U. S. A.  96721 \\
% \texttt{firmjay@hotmail.com} 
 } 

\authorrunning{Firmin J. Oliveira}

\date{Received: 4 October 2013 / Accepted: 4 April 2014}

\maketitle

\begin{abstract}
In this paper the four-dimensional (4-D) space-velocity Cosmological General Relativity of Carmeli
is developed by a general solution of the Einstein field equations.
The Tolman metric is applied in the form
\begin{equation}
  ds^2 = g_{\mu \nu} dx^{\mu} dx^{\nu} = \tau^2 dv^2 -e^{\mu} dr^2
              - R^2 \left( d{\theta}^2 + {\rm sin}^2{\theta} d{\phi}^2 \right),
           \label{abs:tolman-line-element}
\end{equation}
where $g_{\mu \nu}$ is the metric tensor.   We use comoving
coordinates  $x^{\alpha} = (x^0, x^1, x^2, x^3) = (\tau v, r, \theta, \phi)$, where 
$\tau$ is the Hubble-Carmeli time constant, $v$ is the universe expansion velocity
and $r$, $\theta$ and $\phi$ are the spatial coordinates. We assume 
that $\mu$ and $R$ are each functions of the coordinates $\tau v$ and  $r$.

The vacuum mass density $\rho_{\Lambda}$ is defined in terms of a cosmological constant $\Lambda$,
\begin{equation}
    \rho_{\Lambda}  \equiv  -\frac{  \Lambda } { \kappa \tau^2 },   \label{abs:rho_V-def}
\end{equation}
where the Carmeli gravitational coupling constant $\kappa = 8 \pi G / c^2  \tau^2$, where $c$ is the speed of light
in vacuum. This allows the definitions of the effective mass density
\begin{equation}
  \rho_{eff}  \equiv  \rho + \rho_{\Lambda}     \label{abs:rho_eff_fin}
\end{equation}
and effective pressure
\begin{equation}
  p_{eff}  \equiv  p - c \tau  \rho_{\Lambda},     \label{abs:p_eff_fin}
\end{equation}
where $\rho$ is the mass density and $p$ is the pressure.
Then the energy-momentum tensor
\begin{equation}
  T_{\mu \nu} = \tau^2 \left[ \left( \rho_{eff}  +  \frac{p_{eff}} {c \tau} \right) u_{\mu} u_{\nu} -  \frac{p_{eff}} {c \tau} g_{\mu \nu} \right],
        \label{abs:T_uv_eff} 
\end{equation}
where $u_{\mu} = (1, 0, 0, 0)$ is the 4-velocity.
The Einstein field equations are taken in the form
\begin{equation}
   R_{\mu \nu}  = \kappa \left( T_{\mu \nu}  -  \frac{1} {2} g_{\mu \nu}  T \right),
           \label{abs:field-eqs-new}
\end{equation}
where $R_{\mu \nu} $ is the Ricci tensor,
$\kappa = 8 \pi G / c^2 \tau^2$ is Carmeli's gravitation constant, where $G$ is Newton's constant
and the trace $T = g^{\alpha \beta} T_{\alpha \beta}$.
By solving the field equations  (\ref{abs:field-eqs-new}) a space-velocity cosmology is obtained analogous to the
Friedmann-Lema{\^{i}}tre-Robertson-Walker  space-time cosmology. 

We choose an equation
of state such that
\begin{equation}
  p  =  w_e  c \tau \rho,  \label{abs:p-w-rho}
\end{equation}
with an evolving state parameter
\begin{equation}
   w_e \left( R_v \right)  =  w_0  +  \left(1 - R_v \right) w_a,   \label{abs:we-def}
\end{equation}
where $R_v = R_v(v)$ is the scale factor and $w_0$ and $w_a$ are constants. 

Carmeli's 4-D space-velocity cosmology is derived as a special case.


\end{abstract}

\section{Introduction}

Cosmological General Relativity (CGR) is a 5-D time-space-velocity theory\cite{carmeli-0,behar-carmeli} of the cosmos,
for one dimension of cosmic time,  three of space and one of the universe expansion velocity.
Cosmic time is taken to increase from the present epoch $t=0$  toward the big bang time $t= \tau$, where $\tau$ is 
the Hubble-Carmeli time constant.   The expansion velocity $v=0$ at the present epoch and increases toward the big bang. 
In this paper the cosmic time $t$ is held fixed ($dt=0$) and measurements are referred to the present
epoch of cosmic time.  This is a reasonable approach since the time duration over which observations are made is negligible
compared to the travel time of the emitted light from the distant galaxy.

Hence, in this paper we examine the four-dimensional (4-D) space-velocity of CGR.  A general solution to the Einstein field
equations in the space-velocity domain is obtained, analogous to the Friedmann-Lema{\^{i}}tre-Robertson-Walker (FLRW)
solution of space-time cosmology.
The main emphasis herein is to develop a cosmology having a scale factor $R_v$ dependent on the expansion velocity
which in turn can be expressed in terms of the cosmological redshift $z$.  This will enable a 
set of well defined tools for the analysis of observational data where the cosmological redshift plays a central role.
The resulting cosmology is used to model a small set of SNe Ia data.

We will derive Carmeli's cosmology as a special case where the scale factor is held fixed.
Two principal results of Carmeli's cosmology is the prediction of the accelerated expansion of the universe\cite{carmeli-1}
and the description of spiral galaxy rotation curves without additional dark matter\cite{hartnett-2}.  We continue to support 
those results within this paper with a theoretical framework that accommodates a more varied parameter space. 

For our purposes, a vacuum mass density $\rho_{\Lambda}$ is defined in terms of a cosmological constant $\Lambda$ by
\begin{equation}
    \rho_{\Lambda}  \equiv  -\frac{  \Lambda } { \kappa \tau^2 },   \label{eq:rho_V-intro-def}
\end{equation}
where the Carmeli gravitational coupling constant $\kappa = 8 \pi G / c^2  \tau^2$, where $c$ is the speed of light
in vacuum and $\tau$ is the Hubble-Carmeli time constant.  In a previous article\cite{oliveira-1} we hypothesized 
that the observable universe is one of two black holes joined at their event horizons.  From this perspective we show that
the vacuum density of the observable universe and the  universe black hole entropy have the relation
$\rho_{\Lambda} \propto  S^{-1} $, where  $S$ is the Bekenstein-Hawking entropy\cite{bekenstein-1,hawking-1}
of the black hole. We also will use an evolving two parameter equation of state $w_e(R_v)$, dependent on the scale factor
$R_v$, which allows for the evolution of the effect of dark energy on the pressure\cite{fang}.

\section{The Metric}

Assuming the matter in the universe to be isotropically distributed we will adopt a metric that is spatially spherical symmetric.
Furthermore, the spatial coordinates will be co-moving such that galaxies expanding along the same geodesic curve are motionless
with respect to one another.  In this manner we can compare observations between galaxies moving along different geodesic 
paths.

A general derivation of the metric we will use was given by Tolman\cite{tolman-1} and is taken in the simplified form defined by
\begin{equation}
  ds^2 = g_{\mu \nu} dx^{\mu} dx^{\nu} = \tau^2 dv^2 -e^{\mu} dr^2
              - R^2 \left( d{\theta}^2 + {\rm sin}^2{\theta} d{\phi}^2 \right),
           \label{eq:tolman-line-element}
\end{equation}
where $g_{\mu \nu}$ is the metric tensor.
The comoving coordinates are $x^{\alpha} = (x^0, x^1, x^2, x^3) = (\tau v, r, \theta, \phi)$,
 where $\tau$ is the Hubble-Carmeli time constant, $v$ is the universe expansion velocity
and $r$, $\theta$ and $\phi$ are the spatial coordinates. Assume that the functions
 $\mu$ and $R$ are functions of coordinates $\tau v$ and  $r$.

The constant $\tau$ is related to the Hubble constant $H_0$ at zero distance and zero gravity by the 
relation $h = 1 / \tau$ where measurements of  $H_0$ at very close (local) distances
are used to determine the value of $h$.  At this writing the accepted value\cite{oliveira-hartnett-0} is
\begin{equation}
  h  =  72.17 \pm 0.84 \pm 1.64 \, {\rm km / s / Mpc}.    \label{eq:h-value} 
\end{equation}
Therefore
\begin{equation}
  \tau = \left( 4.28 \pm 0.15 \right)   \times 10^{17} {\rm s}  = 13.56 \pm 0.48 \, {\rm Gyr}.  \label{eq:tau-value}
\end{equation}

From  (\ref{eq:tolman-line-element}) the non-zero components of the metric tensor $g_{\mu \nu}$ are given by
\begin{eqnarray}
  g_{0 0} &=& 1,  \label{eq:g_00} \\
 \nonumber \\
  g_{1 1} &=& -e^{\mu},  \label{eq:g_11} \\
 \nonumber \\
  g_{2 2} &=& -R^2,  \label{eq:g_22} \\
 \nonumber \\
  g_{3 3} &=& -R^2 {\rm sin}^2 {\theta}.  \label{eq:g_33} 
\end{eqnarray}
The choice of the particular  metric (\ref{eq:tolman-line-element}) determines that the 4-velocity of a point moving
along the geodesic curve is given by
\begin{equation}
  u_{\mu} = u^{\mu} = dx^{\mu} / ds  = (1, 0, 0, 0).  \label{eq:u_alpha}
\end{equation}

The universe expands by the null condition $ds = 0$.  For a spherically symmetric expansion one has $d{\theta} = d{\phi} = 0$.
The metric (\ref{eq:tolman-line-element}) then gives
\begin{equation}
  \tau^2 dv^2 -e^{\mu} dr^2  = 0, \label{eq:expansion-eq}
\end{equation}
which yields
\begin{equation}
  \frac{dr} {dv} = \tau e^{-\mu / 2}.     \label{eq:dr/dv-0}
\end{equation}

To determine the functions $\mu$ and $R$  we need to solve the Einstein field equations. 

\section{The Field Equations}

The Einstein field equations with a cosmological constant $\Lambda$ term are taken in the form
\begin{equation}
   R_{\mu \nu}  + \Lambda g_{\mu \nu} = \kappa \left( T^{'}_{\mu \nu}  - \frac{1} {2} g_{\mu \nu}  T^{'} \right),   \label{eq:field-eqs-0}
\end{equation}
where $R_{\mu \nu}$ is the Ricci tensor, $T^{'}_{\mu \nu}$  is the energy-momentum tensor,
 $T^{'} = g^{\alpha \beta} T^{'}_{\alpha \beta}$ is its trace and $\kappa$
is Carmeli's gravitational coupling constant given by
\begin{equation}
  \kappa = \frac{8 \pi G} {c^2 \tau^2},   \label{eq:kappa}
\end{equation}
where $G$ is Newton's gravitational constant and $c$ is the speed of light in vacuum.
If we add the tensor $\Lambda g_{\mu \nu}$  to the Ricci tensor $R_{\mu \nu}$,
the covariant derivative of the new tensor is still zero. That is
\begin{equation} 
   \nabla_{\nu} \left( R_{\mu \nu} + \Lambda g_{\mu \nu} \right) =   \nabla_{\nu} R_{\mu \nu}
         + \Lambda  \nabla_{\nu} g_{\mu \nu}  = 0,  \label{eq:Ricci-covariant-dev}
\end{equation}
since the covariant derivatives of the Ricci tensor and the metric tensor are both zero.

We will move the cosmological constant term from the left hand side (l.h.s.) to the right hand side (r.h.s.) of 
 (\ref{eq:field-eqs-0})  to make it a component of the energy-momentum tensor giving
\begin{equation} 
   R_{\mu \nu}  = \kappa \left( T^{'}_{\mu \nu} 
                           - \frac{\Lambda} {\kappa}  g_{\mu \nu} - \frac{1} {2} g_{\mu \nu}  T^{'} \right). 
                \label{eq:field-eqs-1}
\end{equation}

Since the covariant derivative of the energy-momentum tensor $\nabla_{\nu}(T^{'}_{\mu \nu}) = 0$,
 the covariant derivative of the r.h.s. of (\ref{eq:field-eqs-1}) equals zero.
The $\Lambda$ term is absorbed into a new energy-momentum tensor by the form
\begin{equation}
      T_{\mu \nu} = \tau^2 \left[ \left( \rho  + \frac{p} {c \tau}  \right) u_{\mu} u_{\nu}
     - \left( \frac{p} {c \tau} + \frac{\Lambda} {\tau^2 \kappa} \right) g_{\mu \nu} \right],  \label{eq:Tuv_def}
\end{equation}
where $\rho$ is the mass density, $p$ is the pressure and
 $u_{\mu} = u^{\mu} = dx^{\mu} / ds  = (1, 0, 0, 0)$ is the 4-velocity. 
The $\tau^2$ factor on the r.h.s. of (\ref{eq:Tuv_def}) is an artifact of convenience in solving the field equations.

Taking the trace $T$ of the new energy-momentum tensor yields
\begin{eqnarray}
   T &=& g^{\alpha \beta} T_{\alpha \beta} =  \tau^2 \left[  \rho  + \frac{p} {c \tau}
        - 4 \left(   \frac{p} {c \tau} +  \frac{\Lambda} {\tau^2 \kappa} \right)    \right]    \label{eq:T-trace-1}  \\
  \nonumber \\
      &=&  \tau^2  \rho -  3 \frac{ \tau } { c }  p  -  4 \tau^2 \left(  \frac{\Lambda} {\tau^2 \kappa} \right).   \nonumber
\end{eqnarray}

Define the vacuum mass density $\rho_{\Lambda}$ in terms of the cosmological constant $\Lambda$,
\begin{equation}
    \rho_{\Lambda}  \equiv  -\frac{  \Lambda } { \kappa \tau^2 }.   \label{eq:rho_V-def}
\end{equation}
By this definition we are defining $\rho_{\Lambda}$ to be a negative density, unless $\Lambda$ is negative.
This is contrary to the design of the standard model but is in keeping with the Behar-Carmeli cosmological
model\cite{behar-carmeli}.

Then in terms of the vacuum mass density $\rho_{\Lambda}$ , define the effective mass density and effective pressure,
\begin{equation}
  \rho_{eff}  \equiv  \rho + \rho_{\Lambda},     \label{eq:rho_eff_fin}
\end{equation}
and
\begin{equation}
  p_{eff}  \equiv  p - c \tau  \rho_{\Lambda}.     \label{eq:p_eff_fin}
\end{equation}
With the  definitions for the effective mass density and the effective pressure the energy-momentum tensor (\ref{eq:Tuv_def}) becomes
\begin{equation}
  T_{\mu \nu} = \tau^2 \left[ \left( \rho_{eff}  +  \frac{p_{eff}} {c \tau} \right) u_{\mu} u_{\nu} -  \frac{p_{eff}} {c \tau} g_{\mu \nu} \right].
        \label{eq:T_uv_eff} 
\end{equation}
In terms of the effective mass density and pressure the only non-zero components of $T_{\mu \nu}$ are given by
\begin{eqnarray}
  T_{0 0}  &=&  \tau^2 \left( \rho_{eff} + \frac{ p_{eff} } {c \tau } \right) - \tau^2 \left( \frac{ p_{eff} } { c \tau } \right)
                         =  \tau^2 \rho_{eff},    \label{eq:T00-eff}  \\
  \nonumber \\
  T_{1 1}  &=&  \frac{ \tau } { c }  e^{\mu} p_{eff},   \label{eq:T11-eff}  \\
  \nonumber \\
  T_{2 2}  &=&  \frac{ \tau } { c }  R^2 p_{eff},   \label{eq:T22-eff}  \\
  \nonumber \\
  T_{3 3}  &=&  \frac{ \tau } { c }  R^2 {\rm sin}^2 \theta p_{eff}.   \label{eq:T33-eff}
\end{eqnarray}
The trace of $T_{\mu \nu}$ in terms of the effective mass density and pressure is
\begin{equation}
  T = \tau^2 \rho_{eff} - 3  \frac{ \tau } { c } p_{eff}.   \label{eq:T-eff}
\end{equation}
Substituting for the defined values of $\rho_{eff}$ and $p_{eff}$, the trace (\ref{eq:T-eff}) expands out to
\begin{equation}
      T =  \tau^2  \rho -  3 \frac{ \tau } { c }  p  -  4 \tau^2 \left(  \frac{\Lambda} {\tau^2 \kappa} \right),  \label{eq:T-eff-expand}
\end{equation}
which is equal to (\ref{eq:T-trace-1}).
With our definition for the new energy-momentum tensor the field equations now
take the form
\begin{equation}
   R_{\mu \nu}  = \kappa \left( T_{\mu \nu}  -  \frac{1} {2} g_{\mu \nu}  T \right). 
           \label{eq:field-eqs-new}
\end{equation}

Here we write out the nonvanishing components of the Ricci tensor.  A dot $(\,\dot{}\,)$ denotes partial differentiation
with respect to  $x^0 = \tau v$ and a prime $\,(\,')$  denotes partial differentiation with respect to  $x^1 = r$.  This follows\cite{carmeli-0}, (Sect. 4.3.)
\begin{eqnarray}
  R_{0 0}  &=&  -\frac{1}{2} \ddot{\mu} - \frac{2}{R} \ddot{R} - \frac{1}{4} {\dot{\mu}}^2,   \label{eq:R00}  \\
  \nonumber \\
  R_{0 1}  &=&  \frac{1}{R} R' \dot{\mu} - \frac{2}{R} {\dot{R}}',    \label{eq:R01}  \\
  \nonumber \\
  R_{1 1} &=&  e^{\mu} \left( \frac{1}{2} \ddot{\mu} + \frac{1}{4} {\dot{\mu}}^2 
                         + \frac{1}{R} \dot{\mu} \dot{R} \right) + \frac{1}{R} \left( \mu' R' - 2 R'' \right),   \label{eq:R11}  \\
  \nonumber \\
  R_{2 2}  &=&  R \ddot{R} + \frac{1}{2} R \dot{R} \dot{\mu} + {\dot{R}}^2 
                         + 1 - e^{-\mu} \left( R R'' - \frac{1}{2} R R' \mu' + {R'}^2 \right),       \label{eq:R22}  \\
  \nonumber \\
  R_{3 3}  &=&  {\rm sin}^2\theta R_{2 2}.       \label{eq:R33}
\end{eqnarray}

Expanding the r.h.s. of the field equations (\ref{eq:field-eqs-new}) yields
\begin{eqnarray}
   R_{0 0}  &=&  \frac{1}{2} \tau^2 \rho_{eff} + \frac{3}{2} \frac{\tau}{c} p_{eff},    \label{eq:R00=T00-g00T/2}  \\
  \nonumber \\
  R_{0 1}  &=&  0,      \label{eq:R01=T01-g01T/2}  \\
  \nonumber \\
  R_{1 1}  &=&  \frac{1}{2} \tau^2  e^{\mu} \rho_{eff} - \frac{1}{2} \frac{\tau}{c}  e^{\mu} p_{eff},    \label{eq:R11=T11-g11T/2}  \\
  \nonumber \\
  R_{2 2}  &=&  \frac{1}{2} \tau^2  R^2  \rho_{eff}- \frac{1}{2} \frac{\tau}{c}  R^2 p_{eff},   \label{eq:R22=T22-g22T/2}  \\
  \nonumber \\
  R_{3 3}  &=&  \frac{1}{2} \tau^2  R^2 {\rm sin}^2 \theta \rho_{eff} - \frac{1}{2} \frac{\tau}{c}  R^2 {\rm sin}^2 \theta p_{eff}.
                        \label{eq:R33=T33-g33T/2}
\end{eqnarray}

We obtain our first independent field equation by multiplying (\ref{eq:R11=T11-g11T/2}) by $e^{-\mu}$ and adding
the result to (\ref{eq:R00=T00-g00T/2}).  This operation will eliminate the $\ddot{\nu}$ and ${\dot{\mu}}^2$ terms leaving
\begin{equation}
  -\frac{2}{R}\ddot{R} + \frac{1}{R} \dot{\mu} \dot{R} + e^{-\mu} \frac{1}{R} \left( \mu' R' - 2 R'' \right)
          = \kappa \tau^2 \rho_{eff}  +  \kappa \frac{\tau}{c} p_{eff}.   \label{eq:basic-1a}
\end{equation}
By multiplying (\ref{eq:R22=T22-g22T/2}) by $2/R^2$ and adding the result to (\ref{eq:basic-1a}), the $\ddot{R}$ 
and $p_{eff}$ terms will be eliminated leaving
\begin{equation}
  \frac{\dot{\mu} \dot{R}}{R} + \left( \frac{\dot{R}}{R} \right)^2 +\frac{1}{R^2}
         + e^{-\mu} \left[ \frac{ \mu' R'}{R}  - \frac{2 R''}{R}  - \left(\frac{R'}{R} \right)^2 \right]  
            = \kappa \tau^2 \rho_{eff}.   \label{eq:basic-1b}
\end{equation}

The next basic independent field equation is obtained by substituting for the value of the expression $e^{-\mu} (\mu' R' - 2 R'') / R$ in
 (\ref{eq:basic-1b}).  To do that, multiply (\ref{eq:R22=T22-g22T/2}) by $2 / R^2$ and move all other terms to the r.h.s. leaving
\begin{eqnarray}
  e^{-\mu} \left( \frac{\mu' R'}{R} - \frac{2 R''}{R} \right) =   &-& \frac{2 \ddot{R}}{R} - \frac{\dot{R} \dot{\mu}}{R}
                - 2 \left( \frac{\dot{R}}{R} \right)^2 - \frac{2}{R^2}    \label{eq:basic-2a} \\
  \nonumber \\
                           &+& 2 e^{-\mu} \left( \frac{R'}{R} \right)^2
                             + \kappa \tau^2 \rho_{eff} - \kappa \frac{\tau}{c} p_{eff}.    \nonumber  
\end{eqnarray}

Substitute the expression in (\ref{eq:basic-2a}) for its corresponding expression in (\ref{eq:basic-1b}). After eliminating
some terms, combining other terms, multiplying both sides by $-R^2 e^{\mu}$  and then simplifying, we obtain
the next basic field equation
\begin{equation}
  e^{\mu} \left( 2 R \ddot{R} + {\dot{R}}^2 + 1 \right) - {R'}^2  =  -\kappa \frac{\tau}{c}   R^2 e^{\mu} p_{eff}.    \label{eq:basic-2b}
\end{equation}

We restate the last basic field equation we need, which is (\ref{eq:R01=T01-g01T/2}),
\begin{equation}
  2 {\dot{R}}' - R' \dot{\mu} = 0.      \label{eq:basic-3}
\end{equation}

Equations (\ref{eq:basic-1b}),  (\ref{eq:basic-2b}), (\ref{eq:basic-3})  correspond to Carmeli's\cite{carmeli-0} eqns. (4.3.31), (4.3.29) and (4.3.30) 
respectively.

\section{Solutions to the Field Equations}

Equation (\ref{eq:basic-3}) can be partially integrated with respect to $x^0 = \tau v$, keeping $r$ constant.  Integrating it we have
\begin{equation}
  \int^{R'}_{R'_o}{ 2 \frac{\partial{(X')}}{X'} }  = \int^{\mu}_{\mu_o}{ \partial{Y} },  \label{eq:int-1}
\end{equation}
with the result
\begin{equation}
  2 \, {\rm ln}\left( R' / R'_o \right)  = \mu - \mu_o,   \label{eq:int-mu-1a}
\end{equation}
which can be put in the form
\begin{equation}
  \frac{\left( R' \right)^2}{e^{\mu}} = \frac{ \left( R'_o \right)^2}{e^{\mu_o}}  = 1 + f,   \label{eq:int-mu-1b}
\end{equation}
where $f$ is an arbitrary function. The integration constants $R'_o$ and $\mu_o$  are both evaluated at some
particular value of coordinate $x^0 = \tau v_o$  so they can only be  functions of coordinate $r$.  Thus
$f$ is a function of $r$ only.  We can then write (\ref{eq:int-mu-1b}) in the useful form
\begin{equation}
  e^{\mu} =  \frac{ \left( R' \right)^2 }{ 1 + f\left( r \right) }.  \label{eq:int-mu-1c}
\end{equation}
 Since the l.h.s. of 
(\ref{eq:int-mu-1c}) is positive definite,  the r.h.s. is likewise, implying that the condition on $f(r)$ is
\begin{equation}
  1 + f\left( r \right)  >  0.      \label{eq:f(r)-condition} 
\end{equation}
The solution (\ref{eq:int-mu-1c}) is the same as\cite{carmeli-0}, eqn. (4.3.16).

For the  function $R$ we assume the general form
\begin{equation}
  R =   R_r R_v,  \label{eq:R=rRv}
\end{equation}
where
\begin{equation}
  R_r = R_r\left( r \right)  \label{eq:Rr}
\end{equation}
is a function of $r$ only and
\begin{equation}
  R_v = R_v\left( \tau v \right)  \label{eq:Rv}
\end{equation}
is a function of coordinate $\tau v$ only.

Before proceeding with the solution of (\ref{eq:basic-2b}), which has the pressure $p_{eff}$,
we first determine the relevant partial derivatives of (\ref{eq:R=rRv}),
\begin{eqnarray}
  R'  =  R^{'}_r R_v,   \label{eq:R'} \\
  \nonumber \\
  \dot{R} = R_ r \dot{R}_v,   \label{eq:R-dot} \\
  \nonumber \\
  \ddot{R} =  R_r \ddot{R}_v.  \label{eq:R-doubledot}
\end{eqnarray}
Substituting for $e^{\mu}$ and the above derivatives into (\ref{eq:basic-2b}) yields
\begin{eqnarray}
          & & \left( \frac{\left( R^{'}_r \right)^2 R^2_v}{1 + f\left( r \right)} \right)
                   \left[ 2 R^2_r R_v  \ddot{R}_v + R^2_r  \dot{R}^2_v + 1 \right]  - \left( R^{'}_r \right)^2 R^2_v
                                                   \label{eq:basic-2b-sol-1a}  \\
  \nonumber  \\
         &=&    -\kappa \frac{\tau}{c} \left( \frac{\left( R^{'}_r \right)^2  R^2_v} {1 + f \left( r \right) } \right) R^2_r R^2_v p_{eff}.
                                      \nonumber      
\end{eqnarray}
Multiply (\ref{eq:basic-2b-sol-1a}) by $( 1 + f(r) ) / \left( R^{'}_r \right)^2  R^2_v$, gather terms and simplify
to obtain
\begin{equation}
  2 R_v \ddot{R}_v + \left( \dot{R}_v \right)^2  +  \kappa \frac{\tau}{c} R^2_v  p_{eff}  =  \frac{f\left( r \right) }{R^2_r}  =  f_o. 
            \label{eq:basic-2b-sol-1b}
\end{equation}
Assuming the pressure $p_{eff}$ is not a function of $r$, then since the l.h.s. of (\ref{eq:basic-2b-sol-1b}) is a function 
of $x^0 = \tau v$ only and the r.h.s. is a function of $r$ only, they both must equal a constant $f_o$.  From the r.h.s. we conclude that
\begin{equation}
   f\left( r \right) = f_o R^2_r.   \label{eq:f-function-form}
\end{equation}

We next solve (\ref{eq:basic-1b}) which contains the mass density $\rho_{eff}$.  However, we need to first determine
the derivatives of the function $\mu$.  Using (\ref{eq:int-mu-1c}) we obtain
\begin{eqnarray}
  \mu' &=&  \frac{2 R^{''}_r } {R_r}  - \frac{ f' } { 1 + f },   \label{eq:mu-prime}  \\
  \nonumber \\
  \dot{\mu}  &=&  2 \frac{ \dot{R}_v }{ R_v }.   \label{eq:mu-dot}    
\end{eqnarray}
Substituting for the functions and derivatives in (\ref{eq:basic-1b}) we obtain in unsimplified form
\begin{eqnarray}
  \left( \frac{ 1 + f }{ \left( R^{'}_r \right)^2  R^2_v } \right) & & \left[ \left( \frac{ R^{'}_r R_v }{ R_ r R_v } \right)
                                                   \left( \frac{ 2 R^{''}_r } { R_r R_v}  -   \frac{  f' }{ 1 + f } \right)
                                                 - \left(  \frac{ R^{'}_r R_v }{ R_r R_v } \right)^2  - \frac{ 2 R^{''}_r R_v } { R_r R_v } \right] 
                                    \nonumber    \\
  \nonumber \\
                         & &  + \left( \frac{ 2 \dot{R}_v }{ R_v }  \right)  \left( \frac{ R_r \dot{R}_v }{ R_r  R_v }  \right)
                                          +  \left( \frac{ R_r \dot{R}_v }{ R_r R_v }  \right)^2  + \frac{ 1 }{ R^2_r R^2_v }  
                           =   \kappa \tau^2 \rho_{eff}.      \label{eq:basic-1b-sol-1a} 
\end{eqnarray}
This simplifies to
\begin{eqnarray}
& &  3 \left( \dot{R}_v \right)^2   -   \kappa \tau^2 \rho_{eff} R^2_v  =    \label{eq:cal-F-equation}  \\ 
  \nonumber \\
     & &  -\left( \frac{ 1 + f } { \left( R^{'}_r \right)^2 } \right) 
                \left[ \frac{ R^{'}_r } { R_r } \left( \frac{ 2 R^{''}_r } { R^{'}_r R_v } 
                       - \frac{ f' } { 1 + f }  \right) 
                       - \left( \frac{ R^{'}_r } { R_r } \right)^2
                       -  \frac{ 2 R^{''}_r } { R_r } \right] - \frac{ 1 } { R^2_r }   =  {\cal F} \left( v, r \right),   \nonumber
\end{eqnarray}
where in general ${\cal F}$ is a function of $v$ and $r$. 

For the function $R_r$ we now assume the simple form
\begin{equation}
  R_r = r.   \label{eq:R_r-solution}
\end{equation}
Then we have for its derivatives,
\begin{eqnarray}
  R^{'}_r  =  1,   \label{eq:R_r-prime}  \\
  \nonumber \\
  R^{''}_r  =  0.  \label{eq:R_r-doubleprime} 
\end{eqnarray}
As we can see, $R^{''}_r = 0$  implies that its terms along with the $1/R_v$ factor drops out of
(\ref{eq:cal-F-equation}).

Substituting the solution values (\ref{eq:R_r-solution}), (\ref{eq:R_r-prime})  and    (\ref{eq:R_r-doubleprime})
 into the r.h.s. of (\ref{eq:cal-F-equation}) and simplifying we obtain a first order differential equation
of the function $f$ given by
\begin{equation}
  r  f^{'} + f - {\cal F}  r^2 =   r  f^{'} + f - F_o  r^2 =  0,  \label{eq:f-differential} 
\end{equation}
where ${\cal F}(v,r) = F_o$ because the r.h.s. of (\ref{eq:cal-F-equation}) has become a function of $r$ only,
while the l.h.s. is a funcion of $v$ only, so they both must equal the constant $F_o$.
By (\ref{eq:f-function-form}) and (\ref{eq:R_r-solution})  we obtain
\begin{equation}
  f\left( r \right)  = f_o R^2_r  =  f_o r^2,   \label{eq:f(r)-inhomo-sol}
\end{equation}
which implies that, for this solution of $R_r = r$,
\begin{equation}
  F_o = 3 f_o.    \label{eq:F0=3fo}
\end{equation}
If $f_o \ne 0$ then (\ref{eq:f(r)-inhomo-sol}) is the solution to the inhomogeneous differential equation 
(\ref{eq:f-differential}). If $f_o = 0$ then the homogeneous solution is 
\begin{equation}
   f\left( r \right)  =  - \frac{ 2 G M } {c ^2 r },   \label{eq:fr)-homo-sol}
\end{equation}
where the coordinate system is centered on the central mass $M$.
The general solution to (\ref{eq:f-differential}) is the sum of (\ref{eq:f(r)-inhomo-sol}) and (\ref{eq:fr)-homo-sol}),
\begin{equation}
   f\left( r \right)  =  f_o r^2   - \frac{ 2 G M } {c ^2 r }.   \label{eq:fr)-general-sol}
\end{equation}
From (\ref{eq:cal-F-equation}) we have 
\begin{equation}
  \left( \dot{R}_v \right)^2  =  \frac{1}{3} \kappa \tau^2  R^2_v \rho_{eff}  +  f_o.   \label{eq:basic-1b-sol-1b}
\end{equation}

To obtain a value for $f_o$ from (\ref{eq:basic-1b-sol-1b}),
assuming $\dot{R}_v \ne 0$ and $R_v \neq 0$, multiply  the l.h.s. by $R^2_v/R^2_v$
and rearrange the result to obtain
\begin{eqnarray}
  f_o &=&  \frac{1} {c^2} H^2 R^2_v - \frac{1}{3} \kappa \tau^2 R^2_v \rho_{eff}
                      \label{eq:fo-val-1a0}  \\
  \nonumber \\
        &=&  \frac{ H^2 R^2_v } { c^2 } \left( 1  -  \frac{ 8 \pi G } { 3 H^2 } \rho_{eff} \right)  \nonumber  \\
  \nonumber \\
        &=&  \frac{ H^2 R^2_v } { c^2 } \left( 1  -  \frac{ \rho_{eff} } { \rho_{{\cal C}} } \right),  \nonumber
\end{eqnarray}
where $H$ is the Hubble parameter\footnotemark defined by
\begin{equation}
  H \equiv  -c \frac{\dot{R}_v } { R_v },  \label{eq:H_v-def}
\end{equation}
and $\rho_{{\cal C}}$ is the critical density defined by
\begin{equation}
  \rho_{{\cal C}} \equiv \frac{ 3 H^2 } { 8 \pi G }.   \label{eq:rho_C-var-def}
\end{equation}

\footnotetext{To define $H$ as a positive valued parameter in (\ref{eq:H_v-def}) a negative sign is attached to 
compensate for the property that as the expansion velocity $v$ increases from the local frame 
the scale factor $R_v$ decreases toward the origin of the big-bang.  This is contrary to the FLRW definition because there as cosmic time 
$t$ increases from the big-bang to the present the scale factor $a(t)$ also increases. }

Notice that (\ref{eq:fo-val-1a0}) is true at any epoch of coordinate $x^0$, so we will evaluate it at $x^0 = \tau v = 0$.  
We define the  scale factor $R_v$,  the Hubble parameter $H$, the effective mass density $\rho_{eff}$ and the critical density $\rho_{{\cal C}}$ to have the values at the present epoch,
\begin{eqnarray}
  R_v\left( 0 \right)  &=&  1,    \label{eq:Rv0}  \\
  \nonumber \\
  H\left( 0 \right)  &=&  h  =  1 / \tau,  \label{eq:Hv0}  \\
  \nonumber \\
  \rho_{eff}\left( 0 \right)  &=&  \rho_m + \rho_{\Lambda},   \label{eq:rho_effv0}  \\
  \nonumber  \\
   \rho_{{\cal C}}\left( 0 \right)  &=&  \rho_c,   \label{eq:rho_C0}
\end{eqnarray}
where $\rho_m$ is the mass density and $\rho_c$ is the critical density at the present epoch $v=0$.
With the values from (\ref{eq:Rv0})-(\ref{eq:rho_C0}) put into (\ref{eq:fo-val-1a0}) we have
\begin{eqnarray}
  f_o   &=&    \frac{h^2} { c^2 }  \left[ 1   - \frac{ \left( \rho_m  + \rho_{\Lambda} \right) } { \rho_c }  \right]     
            \label{eq:fo-val-1a}   \\
  \nonumber \\
          &=&   \frac{-1} { c^2 \tau^2 }  \left[ \Omega_m   +  \Omega_{\Lambda}  -  1 \right]
            \nonumber  \\ 
 \nonumber \\
          &=&   -K,    \nonumber
\end{eqnarray}
where 
\begin{eqnarray}
  \Omega_m  &=&  \frac{\rho_m } { \rho_c },  \label{eq:Omega_M_def}  \\
  \nonumber \\
  \Omega_{\Lambda}  &=&  \frac{\rho_{\Lambda} } { \rho_c },  \label{eq:Omega_0_def}  \\
  \nonumber \\
  \rho_c  &=&  \frac{ 3 h^2 } { 8 \pi G }   =   \frac{ 3 } { 8 \pi G \tau^2 },    \label{eq:rho_c_def}  \\
  \nonumber \\
  K  &=&  \left(\Omega_m   +  \Omega_{\Lambda}  -  1 \right) / c^2 \tau^2,  \label{eq:K-curvature-1}
\end{eqnarray}
where $\Omega_m$ is the matter mass density parameter at the present epoch ($v=0$),
$\Omega_{\Lambda}$ is the constant vacuum mass density parameter,
and $K$ is the curvature.  Note that the curvature $K$ has the dimension of $[{\rm length}]^{-2}$.
As will be shown,  the type of spatial geometry, hyperbolic (open), Euclidean (flat) or spherical (closed), is determined by the
curvature $K$ which in turn depends on the mass and vacuum densities. This compares with the standard model where 
the type of geometry is determined by the dimensionless curvature parameter $k = -1, 0$ or $1$, for open, flat or closed,
respectively. 

The scale radius $R_0$ is given by 
\begin{equation}
  R_0 = 1 / \sqrt{\mid K \mid}.  \label{eq:scale-radius-R0-def}
\end{equation}
With the value for $f_o$ from (\ref{eq:fo-val-1a}) , with no central mass $M$, $e^{\mu}$ takes the form
\begin{equation}
  e^{\mu}  =  \frac{ R^2_v } { 1 - K r^2 }.  \label{eq:emu}
\end{equation}

\section{Cosmological Redshift  \label{sec:cosmo-redshift} }

For $ds = 0$, (\ref{eq:dr/dv-0}) describes the isotropic expansion of the universe.
Using (\ref{eq:emu})  for $e^{\mu}$, the expansion can be described by
\begin{equation}
  \frac{ dr } { dv }  =  \tau e^{-\mu/2}  =  \frac{ \tau }  { R_v } 
                                   \sqrt{1 -  \left(\Omega_m + \Omega_{\Lambda}  -  1 \right) \left(  r^2 / c^2 \tau^2 \right) }.  \label{eq:dr/dv-1a-00}
\end{equation}
\begin{equation}
  \frac{ \tau dv } { R_v }   =  \frac{ dr } {  \sqrt{ 1 -  K  r^2 }  }.
                           \label{eq:drAr=dvBv-00}
\end{equation}

To show the relation of the scale factor $R_v$ to the cosmological redshift $z$, suppose an observer
in a  galaxy A measures the expansion velocities of two other galaxies ``g'' and ``g+$\delta{\rm g}$'' which is 
nearby galaxy ``g''.
The velocities for the galaxies are $v_A$ and $v_A + \delta{v_A}$ respectively.  Suppose that another observer in
another galaxy O measures expansion velocities for galaxies ``g'' and ``g+$\delta{\rm g}$'' and obtains the values
$v_O$ and $v_O + \delta{v_O}$ respectively. We assume that  $\delta{v_A}$ and $\delta{v_O}$ are small
 compared to  $v_A$ and $v_O$, respectively, since galaxies ``g'' and ``g+$\delta{\rm g}$'' are near each other.
What can we say about the relationship between these measured velocities from galaxies A and O?

Assume the distance to galaxy A is $r_A$ and the distance to galaxy O is $r_{O}$. The galaxies are comoving
which implies that the distance $r_{OA}$ between them is constant.
For the galaxy ``g'', we integrate (\ref{eq:drAr=dvBv-00})
between the points $(v_A, r_A)$ and $(v_O, r_O)$
\begin{equation}
  \int^{v_O}_{v_A}  { \frac{ \tau dv } { R_v } }
      =  \int^{r_O}_{r_A} { \frac{ dr } { \sqrt{ 1 - K r^2 } } }.   \label{eq:int-galax-g}
\end{equation}
For the galaxy ``g+$\delta{\rm g}$'' we integrate  (\ref{eq:drAr=dvBv-00}) between the points
$(v_A + \delta{v_A}, r_A)$  and $(v_O + \delta{v_O},  r_O)$, 
\begin{equation}
  \int^{v_O + \delta{v_O}}_{v_A + \delta{v_A}}  { \frac{ \tau dv } { R_v } }
      =  \int^{r_O}_{r_A} { \frac{ dr } { \sqrt{ 1 - K r^2 } } }.    \label{eq:int-galax-g+dg}
\end{equation}
Subtracting  (\ref{eq:int-galax-g})  from  (\ref{eq:int-galax-g+dg}) we obtain
\begin{eqnarray}
      \int^{v_O + \delta{v_O}}_{v_A + \delta{v_A}}  { \frac{ \tau dv } { R_v } } 
   &-&  \int^{v_O}_{v_A}  { \frac{ \tau dv } { R_v } }  =   \label{eq:int-galax-diff-vel}  \\
  \nonumber \\
  & & \int^{r_O}_{r_A} { \frac{ dr } { \sqrt{ 1 - K r^2 } } }  - \int^{r_O}_{r_A} { \frac{ dr } { \sqrt{ 1 - K r^2 } } }   =  0.   \nonumber
\end{eqnarray}
After some manipulations of the integrals 
(\ref{eq:int-galax-diff-vel}) reduces to
\begin{equation}
   \frac{ \tau \delta{v_O } } { R_v \left( v_O \right) } 
          -   \frac{ \tau \delta{v_A } } {  R_v \left( v_A \right)  }    =     0,   \label{eq:int-delta-vO,delta-vA}
\end{equation}
where $R_v ( v_O )$ is the approximate value of the assumed slowly varying scale factor over the small velocity
interval $\delta{v_O }$, and likewise for $R_v ( v_A )$ over the small interval $\delta{v_A }$.
(\ref{eq:int-delta-vO,delta-vA}) can be put into the form
\begin{equation}\
   \frac{ \tau \delta{v_O } } { \tau \delta{v_A } }  =  \frac{  R_v \left( v_O \right)  } {  R_v \left( v_A \right)  }.  \label{eq:distance-ratio-1}
\end{equation}

If in galaxy A the distance $\tau \delta{v_A}$ is determined by the measurement of $\lambda_A$
of the wavelength  of photons from galaxy ``g + $\delta{g}$'', then in galaxy O the corresponding measurement of the photons from the
same galaxy will have a wavelength $\lambda_O$. The ratio of the two wavelengths is assumed to be given by 
(\ref{eq:distance-ratio-1}),
\begin{equation}
    \frac{ \tau \delta{v_O } } { \tau \delta{v_A } }  =  \frac{  R_v \left( v_O \right)  } {  R_v \left( v_A \right)  }  \label{eq:distance-ratio-2} 
       = \frac{ \lambda_O } { \lambda_A }   =  1 + z,   \label{eq:distance-ratio-3}
\end{equation}
where $z$ is the cosmological redshift of the photon.  If we take galaxy O to be the local galaxy then 
$v_O = 0$ and we set the scale factor of the local galaxy to unity, $R_v \left( v_O \right) = 1$.
 Then setting $v_A = v$ for the velocity of a general galaxy A,
(\ref{eq:distance-ratio-3})  can be put into the familiar form of the scale factor redshift relation,
\begin{equation}
   R_v \left( v \right)   =  \frac{ 1 } { 1 + z },     \label{eq:cosmo-redshift-relation}
\end{equation}
where
\begin{equation}
  1 + z = \frac{\lambda_O} {\lambda_A},  \label{eq:z-lambda}
\end{equation}
where $\lambda_A$ is the  photon wavelength detected by distant galaxy A and $\lambda_O$ is the
wavelength observed in the local galaxy.

\section{Mass Continuity  \label{sec:mass-continuity} }
By substituting the r.h.s. of (\ref{eq:basic-1b-sol-1b}) into its relevant term in (\ref{eq:basic-2b-sol-1b}) and simplifying
we obtain
\begin{equation}
  2 \ddot{R}_v   =  -\left( \frac{1}{3} \kappa \tau^2  \rho_{eff}  +  \kappa \frac{\tau }{ c } p_{eff}  \right) R_v.   \label{eq:R-doubledot-1}
\end{equation}
We can obtain another expression for $2 \ddot{R}_v$ by differentiating (\ref{eq:basic-1b-sol-1b}) with respect to (w.r.t.) $x^0$ which, 
after simplifying gives
\begin{equation}
   2 \ddot{R}_v   =  \frac{1}{3} \kappa \tau^2  \left( \frac{ R^2_v }{ \dot{R}_v } \right)  \dot{\rho}_{eff}
                 +  \frac{2}{3} \kappa \tau^2   R_v  \rho_{eff}.    \label{eq:R-doubledot-2}
\end{equation}
Combining (\ref{eq:R-doubledot-1}) and (\ref{eq:R-doubledot-2}) and simplifying gives us the effective mass
 density continuity equation
\begin{equation}
  \dot{\rho}_{eff}  =  -3 \left( \rho_{eff} + \frac{ p_{eff} }{ c \tau } \right) \left( \frac{ \dot{R}_v } { R_v } \right).
        \label{eq:eff-mass-continuity}
\end{equation}
Simplifying (\ref{eq:eff-mass-continuity}) yields
\begin{equation}
  \dot{\rho}  =  -3 \left( \rho + \frac{ p }{ c \tau } \right) \left( \frac{ \dot{R}_v } { R_v } \right).
        \label{eq:mass-continuity}
\end{equation}
This  mass density continuity equation  (\ref{eq:mass-continuity})  is identical in form to the energy density
 continuity equation of the standard  FLRW model except that here the rate of mass density change
 is w.r.t. expansion velocity $v$ while the standard model rate is w.r.t. cosmic time $t$.

We choose an evolving equation of state parameter\cite{fang} $w_e$ such that the  pressure
$p$ is related to the mass density $\rho$ by
\begin{equation}
  p  =   w_e  c \tau \rho,  \label{eq:p-w-rho}
\end{equation}
where
\begin{equation}
   w_e \left( R_v \right)  =  w_0  +  \left(1 - R_v \right) w_a,   \label{eq:we-def}
\end{equation}
where $w_0$ and $w_a$ are constants.  The second term on the r.h.s. of (\ref{eq:we-def}) represents the
evolution of the equation of state as a function of expansion velocity.  In particular, this functional form allows for a 
state equation to vary from low dark energy influence when the scale factor $R_v$ was small into becoming dominated
by dark energy at the current unity scale factor.
Substituting for $p$ from (\ref{eq:p-w-rho}) into (\ref{eq:mass-continuity}) we obtain
\begin{equation}
  \dot{\rho}  =  -3 \left( 1 +  w_e \right) \rho  \left( \frac{ \dot{R}_v } { R_v } \right).     
     \label{eq:mass-w-cont} 
\end{equation}
(\ref{eq:mass-w-cont}) can be put into the form
\begin{equation}
  \frac{ d \rho } { \rho } 
           =  -3 \left( 1 +  w_0 + \left( 1 - R_v \right) w_a \right)  \frac{ d R_v } { R_v},    \label{eq:drho/rho==dRv/Rv}
\end{equation}
which upon integration yields
\begin{equation}
  \rho  =  \rho_m e^{ -3 w_a \left( 1 - R_v \right)}  R^{-3 \left( 1 + w_0 + w_a \right) }_v.
           \label{eq:rho=rho0*Rv}
\end{equation}
Divide (\ref{eq:rho=rho0*Rv})  by $\rho_c$  to get the mass density parameter
\begin{equation}
  \Omega  =  \Omega_m e^{ -3 w_a \left( 1 - R_v \right)} R^{-3 \left( 1 + w_0 + w_a \right) }_v.
           \label{eq:omega=omega_m*R_v}
\end{equation}
Converting $R_v$ in terms of the redshift $z$ from (\ref{eq:cosmo-redshift-relation}),
(\ref{eq:omega=omega_m*R_v}) for $\Omega = \Omega(z)$ becomes
\begin{equation}
   \Omega \left( z \right)  =  \Omega_m e^{-3 w_a z / \left( 1 + z \right) } 
                    \left( 1 + z \right)^{3 \left( 1 + w_0 + w_a \right) }.
           \label{eq:omega=omega_m*z}
\end{equation}

\section{Acceleration of the Scale Factor }

With the equation of state (\ref{eq:p-w-rho}) the effective pressure becomes
\begin{equation}
   p_{eff}  =   p - c \tau \rho_{\Lambda}  =  c \tau \left( w_e \rho  -  \rho_{\Lambda} \right).
         \label{eq:eqn-of-state-1}
\end{equation}
By substituting the r.h.s. of (\ref{eq:basic-1b-sol-1b}) into its relevant term in (\ref{eq:basic-2b-sol-1b}) and simplifying
we obtain
\begin{equation}
  2 \ddot{R}_v   =  -\left( \frac{1}{3} \kappa \tau^2  \rho_{eff}  +  \kappa \frac{\tau }{ c } p_{eff}  \right) R_v.   \label{eq:R-doubledot-3}
\end{equation}
Substitute for effective pressure from (\ref{eq:eqn-of-state-1})
into (\ref{eq:R-doubledot-3}) and simplify to obtain the scale factor acceleration equation
\begin{equation}
    2 \ddot{R}_v   =  - \frac{ \kappa \tau^2 } { 3 }  \left[ \left( 1 + 3 w_e \right)  \rho  - 2 \rho_{\Lambda} \right] R_v.
         \label{eq:R-doubledot-3-0}
\end{equation}
This can be put into the form
\begin{equation}
        2 c^2 \tau^2 \ddot{R}_v   =  - \left( 1 + 3 w_e \right) \Omega   R_v  + 2 \Omega_{\Lambda} R_v, 
                                \label{eq:R-doubledot-3-2}
\end{equation}
where $\Omega = \Omega(v)$ is given by (\ref{eq:omega=omega_m*R_v}) 
and  $\Omega_{\Lambda}$ is the vacuum mass density parameter.
(\ref{eq:R-doubledot-3-2})  exhibits a range of possible senarios for accelerating and decelerating expansions
depending on $w_e$ and on the values for $\Omega_m$  and $\Omega_{\Lambda}$.

It was discovered experimentally that the expanding universe makes a transition from accelerating to
decelerating\cite{oliveira-hartnett-0,riess-1,riess-2}.
Assume that the transition took place at velocity $v_t$ corresponding to a redshift $z_t$. 
Then, taking (\ref{eq:R-doubledot-3-2}) and setting $\ddot{R}_v =0$ at $R_v(v_t) = ( 1 + z_t )^{-1}$
we have an expression for $w_e$ in terms of the transition redshift $z_t$,
\begin{equation}
   \left[ 1 + 3 w_0 + 3 w_a z_t  \left( 1 + z_t \right)^{-1} \right] \Omega_{z_t}  - 2 \Omega_{\Lambda}  = 0, 
             \label{eq:w_e-formula}
\end{equation}
where $\Omega_{z_t} = \Omega( z_t )$.  This expression can be used to obtain a value
for $z_t$ in terms of the fitted parameters $\Omega_m$, $\Omega_{\Lambda}$, $w_0$ and $w_a$.
We will encounter the transition redshift in the section on modeling.

\section{Entropy of the Black Hole Universe \label{sec:black-hole-entropy} }

The scale radius of the universe given by (\ref{eq:scale-radius-R0-def})  can be put into the form
\begin{equation}
  R_0 = c \tau  / \sqrt{\mid \Omega_m + \Omega_{\Lambda} - 1 \mid}.  \label{eq:R0}
\end{equation}
The value of $\Omega_m$ is the mass density at the present epoch of cosmic time $t=0$, recalling that in CGR the cosmic time 
is measured from the present time $t=0$ increasing toward the big bang time $t=\tau$.
Assuming the universe is a black hole\cite{oliveira-1} of radius $R_0$, then the event horizon surface area $A$ is given by
\begin{equation}
  A = 4 \pi R^2_0  = \frac{ 4 \pi  c^2 \tau^2 } { \mid \Omega_m + \Omega_{\Lambda} - 1 \mid } .   \label{eq:critical-BH-Area}
\end{equation}
Then the Bekenstein-Hawking\cite{bekenstein-1,hawking-1} entropy $S$ of the black hole universe is given by
\begin{equation}
  S = \frac{ k c^3 A } {4 \hbar G} 
     = \frac{ \pi k c^5 \tau^2   }
               {\hbar G  \mid \Omega_m  + \Omega_{\Lambda} - 1 \mid }, 
                   \label{eq:S-entropy-critical-mass}
\end{equation}
where $k$ is Boltzmann's constant and $\hbar$ is Planck's constant over $2 \pi$.
Multiplying the r.h.s. of (\ref{eq:S-entropy-critical-mass})  by $\rho_c / \rho_c$ and simplifying we obtain
\begin{equation}
   \rho_{\Lambda}  =  \frac{ \rho_P } { S / k  }  + \rho_c - \rho_m,  
                                   \label{eq:vacuum-density-BH-entropy}
\end{equation}
where $\rho_{\Lambda} = \rho_c \Omega_{\Lambda}$,  $\rho_m = \rho_c  \Omega_m$ 
and $\rho_P$ is the cosmological Planck mass density defined by
\begin{equation}
     \rho_P = \frac{ \pm {\cal M}_P } { {\cal L}^3_ P } = \frac{ \pm 3 c^5  } {  8 \hbar G^2 },  \label{eq:Planck_vac_dens}
\end{equation}
where
\begin{equation}
   {\cal M}_P =  \sqrt{ \sqrt{\frac{3}{8}} \frac{\hbar c } {G}}   \label{eq:Planck_mass}
\end{equation}
is the cosmological Planck mass and
\begin{equation}
  {\cal L}_P = \hbar / {\cal M}_P  c,     \label{eq:Planck_length}
\end{equation}
is the cosmological Planck length.
By Eqs. (\ref{eq:S-entropy-critical-mass}-\ref{eq:Planck_length}),  since $\rho_c - \rho_m > 0$ observationally,
and since the entropy is always non-negative, then a positive vacuum mass density implies a positive cosmological 
Planck mass density and visa-versa for a negative vacuum density. A deeper analysis into the relation between black 
hole entropy and the vacuum mass density is beyond the scope of this paper.

\section{Expansion of Universe}

(\ref{eq:basic-1b-sol-1b}) can be put into the form
\begin{eqnarray}
  \dot{R}_v  &=&  \frac{ d{R_v} } { \tau dv }   
                    =   \frac{-1} {c \tau }  \sqrt{ \Omega_{eff} R^2_v  - K c^2 \tau^2 }  \label{eq:dRv/dv}   \\
  \nonumber \\
                    &=&  \frac{-1} {c \tau }  \sqrt{ \left( \Omega  + \Omega_{\Lambda} 
                               \right) R^2_v  - \Omega_K },  \nonumber 
\end{eqnarray}
where we used (\ref{eq:rho_eff_fin}) for $\Omega_{eff}$, $\Omega = \Omega(v)$ is given by  (\ref{eq:omega=omega_m*R_v}) and 
\begin{equation}
  \Omega_K = K c^2 \tau^2 = \left( \Omega_m + \Omega_{\Lambda} - 1 \right)   \label{eq:Omega_K-def}
\end{equation}
is the curvature density parameter.
We select the minus sign in the r.h.s. of (\ref{eq:dRv/dv}) because $R_v$ is 
assumed to be a decreasing function of $v$.

(\ref{eq:dr/dv-0}) describes the isotropic expansion of the universe.  Using (\ref{eq:emu})  for $e^{\mu}$,
 the expansion can be described by
\begin{equation}
  \frac{ \tau dv } { R_v }   =  \frac{ dr } {  \sqrt{1 -  K r^2 }  }.
                           \label{eq:drAr=dvBv}
\end{equation}
From (\ref{eq:dRv/dv}) we obtain
\begin{equation}
  \tau dv   =    \frac{ dR_v } { \dot{R}_v }  =  \frac{ -c \tau \, d{R_v} }
            { \sqrt{ \left( \Omega + \Omega_{\Lambda} \right)  R^2_v  - \Omega_K  }  }.    \label{eq:dv=dRv/C}
\end{equation}
Substituting for $\tau dv$ from (\ref{eq:dv=dRv/C}) into the l.h.s. of (\ref{eq:drAr=dvBv}) and simplifying we derive
\begin{equation}
 \frac{ -c \tau \, d{R_v} }
            { \sqrt{  \left( \Omega + \Omega_{\Lambda}  \right)  R^4_v
                                  - \Omega_ K  R^2_v }   }
              =  \frac{ dr } {  \sqrt{1 -  K r^2 }  }.      \label{eq:dvBv=drAr}
\end{equation}
From (\ref{eq:cosmo-redshift-relation})  the scale factor $R_v$ is related to the cosmological redshift $z$ by
\begin{equation}
   R_v  =  \left( 1 + z \right)^{-1}.    \label{eq:Rv-z}
\end{equation}
Differentiating (\ref{eq:Rv-z}) w.r.t. $z$ gives
\begin{equation}
  d{R_v} = -\left( 1 + z \right)^{-2} dz.    \label{eq:dRv/dz}
\end{equation}
Substituting for $R_v$ and $d{R_v}$ from (\ref{eq:Rv-z}) and  (\ref{eq:dRv/dz}) into  (\ref{eq:dvBv=drAr}) 
and simplifying yields the differential for the comoving distance relation
\begin{equation}
  \frac{ c \tau dz } { \sqrt{ \Omega + \Omega_{\Lambda}
                      - \Omega_K  \left( 1 + z \right)^2 }  }
             =  \frac{ dr } {  \sqrt{1 -  K r^2 }  },     \label{eq:dzBz=drAr}
\end{equation}
where $\Omega = \Omega(z)$ is given by (\ref{eq:omega=omega_m*z}).  The spatial geometry defined by the r.h.s. of
(\ref{eq:dzBz=drAr}) is either hyperbolic (open), Euclidean (flat) or spherical (closed) depending on
curvature $K < 0$, $K=0$ or $K>0$, respectively.
Since $K$ is dependent on the mass and vacuum densities then the geometry of the universe is determined by
$\Omega_m$ and $\Omega_{\Lambda}$.

The expansion defined by (\ref{eq:drAr=dvBv}) when substituted for the l.h.s. of (\ref{eq:dzBz=drAr})
and combined with (\ref{eq:Rv-z})  yields the differential equation for the expansion velocity
\begin{equation}
  dv  =  \frac{ c \, dz } { \sqrt{ \left(  \Omega
                                                   + \Omega_{\Lambda} \right) \left(1 + z \right)^2
                      - \Omega_K  \left( 1 + z \right)^4 }  }.  \label{eq:dv/c}
\end{equation} 
Integrating (\ref{eq:dv/c}) we get for the expansion velocity as a function of redshift
\begin{equation}
  v(z)  =  \int^z_0 {\frac{ c \, dz^{'} } { \sqrt{ \left(  \Omega
                                                   + \Omega_{\Lambda} \right) \left(1 + z^{'} \right)^2
                      - \Omega_K  \left( 1 + z^{'} \right)^4 }  } }.  \label{eq:v_expansion}
\end{equation}

\section{Distances}

The Hubble parameter defined by (\ref{eq:H_v-def}), using (\ref{eq:dRv/dv}), is expressed
in terms of velocity $v$,
\begin{equation}
   H\left(v  \right) = -c \frac{ \dot{R}_v } { R_v }  =   h \sqrt{ \Omega  + \Omega_{\Lambda} 
                  -   \Omega_K R^{-2}_v},   \label{eq:H-func-of-v}
\end{equation}
where $h = 1 / \tau$, $\Omega = \Omega(v)$.
In terms of redshift, using (\ref{eq:Rv-z}) for $R_v$,
\begin{equation}
   H \left( z \right)  =   h \sqrt{ \Omega  + \Omega_{\Lambda}  - \Omega_K \left( 1 + z \right)^2 }, 
         \label{eq:H-func-of-z}
\end{equation}
where $\Omega = \Omega(z)$ is defined by (\ref{eq:omega=omega_m*z}).

 \subsection{Comoving Distance}

The comoving distance $D_C$ is the integral of (\ref{eq:dzBz=drAr}) and can be written in terms of the 
Hubble parameter using (\ref{eq:H-func-of-z}),
\begin{equation}
  D_C  =  c \tau \int^z_0 { \frac{ h dz' } { H \left( z' \right) } } 
          =  \int^r_0 { \frac{ dr' } {  \sqrt{1 -  \Omega_K  r'^2 / c^2 \tau^2 }  } }. 
       \label{eq:comoving_distance}
\end{equation}

\subsection{Transverse Comoving Distance}

The transverse comoving distance $D_M$ is the coordinate distance $r$ which is obtained from
the inversion of (\ref{eq:comoving_distance}), i.e., $D_M = r$. It takes the form
\begin{eqnarray}
   D_M   &=&  \frac{ c \tau } { \sqrt{  -\Omega_K } }
                {\rm sinh} \left(  \frac{\sqrt{ -\Omega_K }}{c \tau} D_C \right) 
                              \; \; {\rm for } \; \Omega_K < 0,  \label{eq:D_M_def}  \\
  \nonumber  \\
   D_M   &=&   c \tau  D_C  \; \; {\rm for } \; \Omega_K = 0 \; \; {\rm and}   \nonumber  \\
  \nonumber  \\
   D_M   &=&   \frac{ c \tau } { \sqrt{ \Omega_K  } }
                {\rm sin} \left(  \frac{\sqrt{ \Omega_K  }}{c \tau} D_C \right) 
                              \; \; {\rm for } \; \Omega_K > 0.  \nonumber
\end{eqnarray}

\subsection{Angular Diameter Distance}

A physical source of size $\Delta{S}$ subtends an  observed angle of $\Delta{\theta}$ on the sky given by the
relation
\begin{equation}
  \Delta{\theta}   =  \frac{ \Delta{S} } { R },   \label{eq:delta_theta}
\end{equation} 
where $R=R_v r$ is the proper distance from the coordinate system origin to the source.
Since $R_v = 1/(1+z)$  is the scale factor and $r=D_M$ is the coordinate distance, we define the
angular diameter distance,
\begin{equation}
  D_A  =  R_v r  =  \frac{ D_M } { 1 + z }.  \label{eq:D_A_def}
\end{equation}

\subsection{Luminosity Distance}

In can be shown\footnotemark  that the source luminosity $L$ transforms due to the universe expansion
 as $1/( 1 + z )^4$. Then the flux $S$ from a source of luminosity $L$ at proper distance $R = R_v r$
 is given by
\begin{equation}
   S  =  \left( \frac{ L } { \left( 1 + z \right)^4 } \right) \left( \frac{1} { 4 \pi R^2_v r^2 }  \right) = 
            \frac{ L } { 4 \pi  \left( 1 + z \right)^2 r^2 },
                      \label{eq:S_bolo_flux_def1}
\end{equation}
where we used $R_v = 1 / 1 + z$.
The bolometric flux $S$ can also be defined for a source of luminosity $L$ at distance $D_L$ by
\begin{equation}
   S  =  \frac{ L } { 4 \pi D^2_L } .  \label{eq:S_bolo_flux_def2}
\end{equation}
Eliminating $S$ between (\ref{eq:S_bolo_flux_def1}) and (\ref{eq:S_bolo_flux_def2}) we have the luminosity distance
in terms of the redshift
\begin{eqnarray}
  D_L \left( z \right) &=&  \left( 1 + z \right) \frac{ c \tau } { \sqrt{ -\Omega_K } }
                {\rm sinh} \left( \sqrt{ -\Omega_K } \int^z_0 { \frac{ h dz' } { H \left( z' \right) } }  \right) 
                              \; \; {\rm for } \; \Omega_K < 0,  \label{eq:D_L_z-def}  \\
  \nonumber  \\
  D_L \left( z \right) &=&    \left( 1 + z \right)  c \tau 
              \int^z_0 { \frac{h  dz' } { H \left( z' \right) } }  \; \; {\rm for } \; \Omega_K = 0 \; \; {\rm and}   \nonumber  \\
  \nonumber  \\
  D_L \left( z \right) &=&   \left( 1 + z \right)  \frac{ c \tau } { \sqrt{ \Omega_K  } }
                {\rm sin} \left( \sqrt{ \Omega_K  } \int^z_0 { \frac{ h dz' } { H \left( z' \right) } }  \right) 
                              \; \; {\rm for } \; \Omega_K > 0,   \nonumber
\end{eqnarray}
where we also used the l.h.s. of (\ref{eq:comoving_distance}),  (\ref{eq:D_M_def}) and the
Hubble parameter defined by (\ref{eq:H-func-of-z}),
\begin{equation}
   H \left( z \right)  =   h \sqrt{ \Omega  + \Omega_{\Lambda}  - \Omega_K \left( 1 + z \right)^2 }, 
         \label{eq:H-func-of-z-repeat}
\end{equation}
where $\Omega = \Omega(z)$ from (\ref{eq:omega=omega_m*z}), 
\begin{equation}
   \Omega \left( z \right)  =  \Omega_m e^{-3 w_a z / \left( 1 + z \right) } 
                    \left( 1 + z \right)^{3 \left( 1 + w_0 + w_a \right) }.
           \label{eq:omega=omega_m*z-repeat}
\end{equation}

\footnotetext{Essentially, due to the expansion, the radiation density decreases by a factor $1/(1+z)^3$
 due to the increase in volume and each photon energy decreases by a factor $1/(1+z)$ due to the 
cosmological redshift.}

\section{Model Applications}

We give a view of the cosmology by applying it to a small combined set\cite{riess-2,astier}
total of 157 high redshift SNe Ia data, distance moduli and errors ($\mu_B \pm \sigma_B$)
but not systematic errors. Since we are expecting a scale factor transition from accelerated to decelerated
expansion at low redshift we require that at the origin the scale factor acceleration $\ddot{R}_v (z=0) > 0$. 
The standard distance modulus relation $m(z)$ is given by
\begin{equation}
   m \left( z \right) - M_B = 5 \log( D_L (z)) + 25  + a,  \label{eq:m(z)-theoretical}
\end{equation}
where $M_B$ is the absolute magnitude of a standard supernova at the peak of its light-curve, 
$D_L(z)$ is the luminosity distance (\ref{eq:D_L_z-def}) and $a$ is an arbitrary zero
point offset\cite{hartnett-oliveira}.   For all our examples, the set of parameters are pre-selected
by trial and error and then a final fit is made of the distance modulus relation to the
data varying only the offset parameter $a$.

For the first example we take for the mass density parameter
\begin{equation}
  \Omega_m = \Omega_b / h^2_0,    \label{eq:omega_m=omega_b/h2}
\end{equation} 
 where  $\Omega_b$ is the baryon density parameter,
and $h_0 = h / 100 \, {\rm km / s / Mpc}$.  From (\ref{eq:h-value}) this gives
$h_0 = 0.7217$.   We use a value of $\Omega_b = 0.020$ from\cite{burles-1}.  Therefore,
our value for the mass density parameter is 
\begin{equation}
  \Omega_m = 0.038.   \label{eq:omega_m-Ex-1}
\end{equation}
By a few trials we found a good fit for a
value of the vacuum density parameter
\begin{equation}
  \Omega_{\Lambda} = -0.019.  \label{eq:omega_lambda-Ex-1}
\end{equation}
This defines an open universe since the curvature 
$K = (\Omega_m + \Omega_{\Lambda} - 1 ) / c^2 \tau^2 = -0.981 / c^2 \tau^2 < 0$. 
We consider an evolving dark energy state with $w_0=-1.0$ and $w_a=+1.0$.

\begin{figure}[htb!]
%\centering %
\setlength{\fboxrule}{1.5pt}
%%%\fbox{\includegraphics[viewport=0 530 175 660,keepaspectratio,clip=true]{Rvdd-DE-2B.eps}}  % bb=120 0 494 694  %
%%%\fbox{\includegraphics[viewport=110 530 285 660,keepaspectratio,clip=true]{Rvdd-DE-2B-new.eps}}  % bb=120 0 494 694  %
%-->\fbox{\includegraphics[viewport=110 530 285 660,keepaspectratio,clip=true]{Rvdd-DE-6A.eps}}  % bb=120 0 494 694  %
\fbox{\includegraphics[viewport=0 530 175 660,keepaspectratio,clip=true]{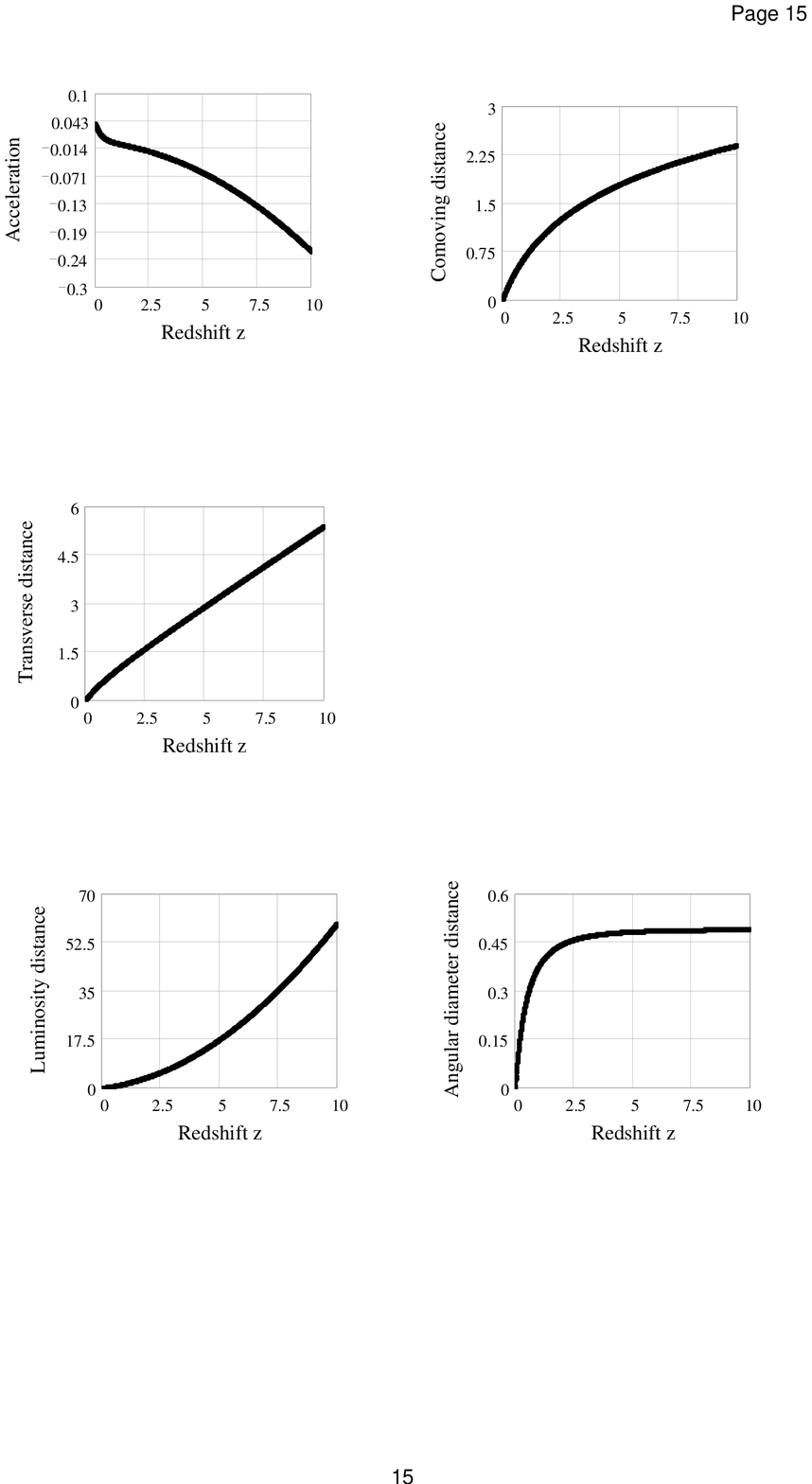}}  % bb=120 0 494 694  %
\caption{Scale factor acceleration $\ddot{R}_v$.
$\Omega_m = 0.038$,  $\Omega_{\Lambda} = -0.019$, $w_0 = -1.0$ and $w_a = 1.0$. }
\label{fig:scalfac-accel}
\end{figure}

\begin{figure}[htb!]
%\centering %
\setlength{\fboxrule}{1.5pt}
%%%\fbox{\includegraphics[viewport=0 460 175 590,keepaspectratio,clip=true]{Rvdd-DE-2A-HST.eps}}  % bb=108 0 508 585  %
%%%\fbox{\includegraphics[viewport=98 460 273 590,keepaspectratio,clip=true]{Rvdd-DE-2A-HST-new.eps}}  % bb=108 0 508 585  %
%-->\fbox{\includegraphics[viewport=98 460 273 590,keepaspectratio,clip=true]{Rvdd-DE-6B2.eps}}  % bb=108 0 508 585  %
\fbox{\includegraphics[viewport=0 460 175 590,keepaspectratio,clip=true]{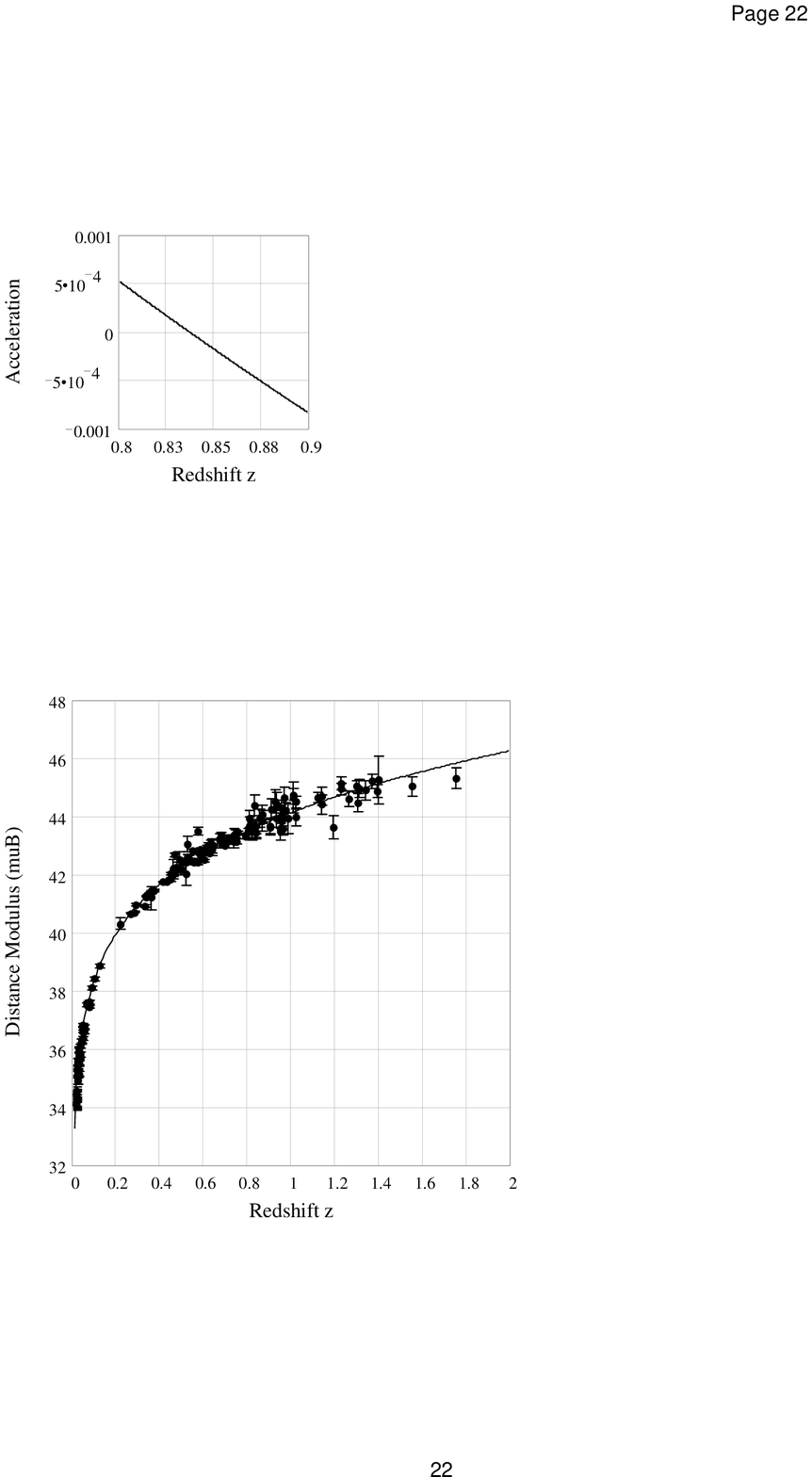}}  % bb=108 0 508 585  %
\caption{Scale factor acceleration $\ddot{R}_v$ at transition
redshift $z_t \approx 0.84$. $\Omega_m = 0.038$, $\Omega_{\Lambda} = -0.019$, 
$w_0 = -1.0$ and $w_a = 1.0$. }
\label{fig:scalfac-accel-trans-z}
\end{figure}

\begin{figure}[htb!]
%\centering %
\setlength{\fboxrule}{1.5pt}
%%%\fbox{\includegraphics[viewport=0 130 215 390,keepaspectratio,clip=true]{Rvdd-DE-2A-HST.eps}}  % bb=108 0 508 585  %
%%%\fbox{\includegraphics[viewport=98 105 313 390,keepaspectratio,clip=true]{Rvdd-DE-2A-HST-new.eps}}  % bb=108 0 508 585  %
%-->\fbox{\includegraphics[viewport=98 105 350 390,keepaspectratio,clip=true]{Rvdd-DE-6B2.eps}}  % bb=108 0 508 585  %
\fbox{\includegraphics[viewport=0 105 252 390,keepaspectratio,clip=true]{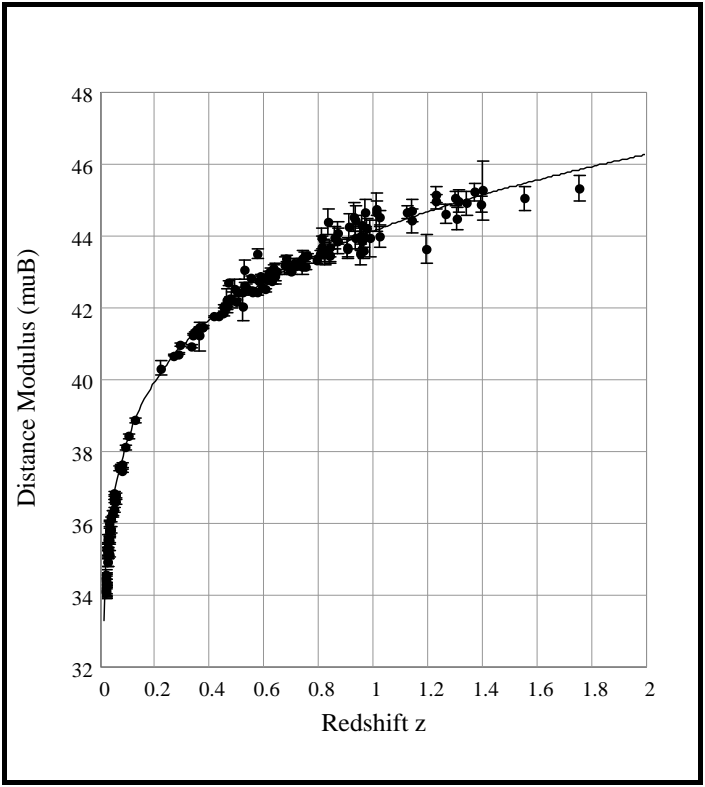}}  % bb=108 0 508 585  %
\caption{SNe Ia distance moduli. $\Omega_m = 0.038$, $\Omega_{\Lambda} = -0.019$,
$w_0 = -1.0$ and $w_a = 1.0$. Points are ($\mu_{\rm B} \pm \sigma_B$) from\cite{riess-2,astier}.
Solid curve is  $m(z) - {\rm M_B} = 5 \log( c \tau D_L (z)) + 25 + a$ with fitted $a = 0.164$ and $\chi^2/157 = 8.809$.}
\label{fig:muB-vs-z}
\end{figure}

Fig. \ref{fig:scalfac-accel} shows that the acceleration $\ddot{R}_v$ starts out positive at $z=0$
and transitions to negative around $z=0.84$. The detail at the acceleration transition is displayed
by Fig. \ref{fig:scalfac-accel-trans-z}.
In Fig. \ref{fig:muB-vs-z} is the plot of the distance moduli for the SNe Ia data.
The theoretical distance modulus (\ref{eq:m(z)-theoretical})  is shown by the solid line.
We obtained a fitted value $a =0.164$ with  $\chi^2/157 = 8.809$.

\begin{figure}[htb!]
%\centering %
\setlength{\fboxrule}{1.5pt}
%%%\fbox{\includegraphics[viewport=5 340 180 470,keepaspectratio,clip=true]{Rvdd-DE-2B.eps}}  % bb=120 0 494 694  %
%%%\fbox{\includegraphics[viewport=109 340 284 470,keepaspectratio,clip=true]{Rvdd-DE-2B-new.eps}}  % bb=120 0 494 694  %
%-->\fbox{\includegraphics[viewport=109 340 284 470,keepaspectratio,clip=true]{Rvdd-DE-6A.eps}}  % bb=120 0 494 694  %
\fbox{\includegraphics[viewport=5 340 180 470,keepaspectratio,clip=true]{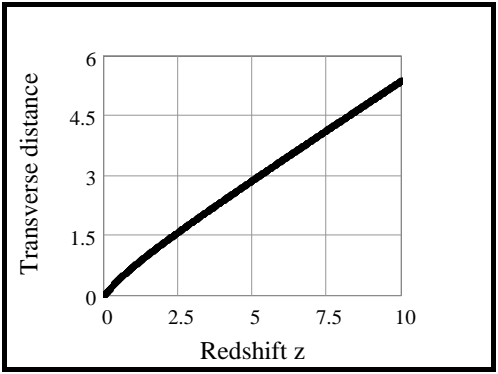}}  % bb=120 0 494 694  %
\caption{Transverse comoving distance $D_M$.  $\Omega_m = 0.038$,
$\Omega_{\Lambda} = -0.019$, $w_0 = -1.0$ and $w_a = 1.0$. }
\label{fig:trnsv-dist}
\end{figure}

\begin{figure}[htb!]
%\centering %
\setlength{\fboxrule}{1.5pt}
%%%\fbox{\includegraphics[viewport=195 525 370 655,keepaspectratio,clip=true]{Rvdd-DE-2B.eps}}  % bb= 120 0 494 694  %
%%%\fbox{\includegraphics[viewport=303 525 478 655,keepaspectratio,clip=true]{Rvdd-DE-2B-new.eps}}  % bb= 120 0 494 694  %
%-->\fbox{\includegraphics[viewport=303 525 478 655,keepaspectratio,clip=true]{Rvdd-DE-6A.eps}}  % bb= 120 0 494 694  %
\fbox{\includegraphics[viewport=195 525 370 655,keepaspectratio,clip=true]{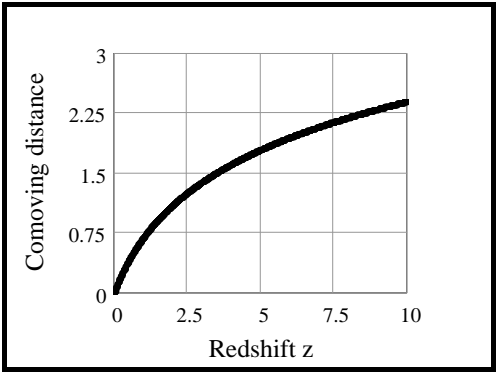}}  % bb= 120 0 494 694  %
\caption{Comoving distance $D_C$.  $\Omega_m = 0.038$,
 $\Omega_{\Lambda} = -0.019$, $w_0 = -1.0$ and $w_a = 1.0$. }
\label{fig:comov-dist}
\end{figure}

\begin{figure}[htb!]
%\centering %
\setlength{\fboxrule}{1.5pt}
%%%\fbox{\includegraphics[viewport=9 160 184 290,keepaspectratio,clip=true]{Rvdd-DE-2B.eps}}  % bb=120 0 494 694  %
%%%\fbox{\includegraphics[viewport=117 160 292 290,keepaspectratio,clip=true]{Rvdd-DE-2B-new.eps}}  % bb=120 0 494 694  %
%-->\fbox{\includegraphics[viewport=117 160 292 290,keepaspectratio,clip=true]{Rvdd-DE-6A.eps}}  % bb=120 0 494 694  %
\fbox{\includegraphics[viewport=9 160 184 290,keepaspectratio,clip=true]{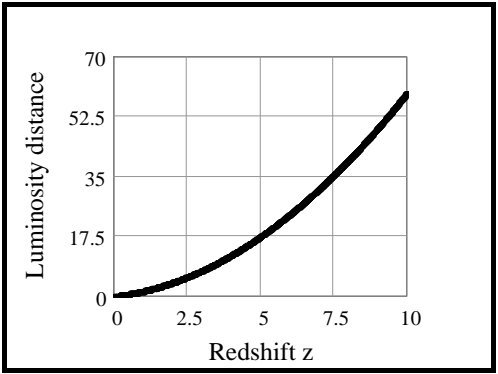}}  % bb=120 0 494 694  %
\caption{Luminosity distance $D_L$.  $\Omega_m = 0.038$, $\Omega_{\Lambda} = -0.019$, $w_0 = -1.0$ and $w_a = 1.0$.}
\label{fig:lum-dist}
\end{figure}

\begin{figure}[htb!]
%\centering %
\setlength{\fboxrule}{1.5pt}
%%%\fbox{\includegraphics[viewport= 202 160 377 290,keepaspectratio,clip=true]{Rvdd-DE-2B.eps}}  % bb=120 0 494 694  %
%%%\fbox{\includegraphics[viewport= 310 160 485 290,keepaspectratio,clip=true]{Rvdd-DE-2B-new.eps}}  % bb=120 0 494 694  %
%-->\fbox{\includegraphics[viewport= 310 160 485 290,keepaspectratio,clip=true]{Rvdd-DE-6A.eps}}  % bb=120 0 494 694  %
\fbox{\includegraphics[viewport= 202 160 377 290,keepaspectratio,clip=true]{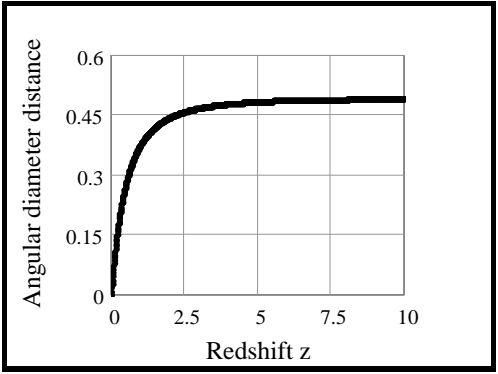}}  % bb=120 0 494 694  %
\caption{Angular diameter distance $D_A$.  $\Omega_m = 0.038$,  $\Omega_{\Lambda} = -0.019$, $w_0 = -1.0$ and $w_a = 1.0$. }
\label{fig:angdiam-dist}
\end{figure}

Figs. \ref{fig:trnsv-dist}-\ref{fig:angdiam-dist} display the various distance relations
all with the same parameters from this first example.  Distances are in units of
 $c \tau = 4158 \, {\rm Mpc}$.

For the second example we select no dark matter
\begin{equation}
  \Omega_m = 0.038,   \label{eq:omega_m-Ex-2}
\end{equation}
with positive vacuum density
\begin{equation}
  \Omega_{\Lambda} = 0.4.  \label{eq:omega_lambda-Ex-2}
\end{equation}
This again defines an open universe since $K = -0.562 / c^2 \tau^2 < 0$.
The equation of state parameters are $w_0 = -1$ and $w_a=+4.3$.

For the third examle we select baryonic and dark matter
\begin{equation}
  \Omega_m = 0.3,   \label{eq:omega_m-Ex-3}
\end{equation}
with positive vacuum density
\begin{equation}
  \Omega_{\Lambda} = 0.7, \label{eq:omega_lambda-Ex-3}
\end{equation}
where
$\Omega_m + \Omega_{\Lambda} = 1$ which gives a flat space curvature $K = 0$ 
as in the standard FLRW Lambda-Cold-Dark-Matter (LCDM) model. We use the same equation
of state $w_0=-1$ and $w_a=+4.3$ as the second example. The results of these two examples are combined in the plots.

\begin{figure}[htb!]
%\centering %
\setlength{\fboxrule}{1.5pt}
%%%\fbox{\includegraphics[viewport=0 460 175 590,keepaspectratio,clip=true]{Rvdd-NoDE1.eps}}  % bb=108 0 508 585  %
%%%\fbox{\includegraphics[viewport=98 460 273 590,keepaspectratio,clip=true]{Rvdd-NoDE1-new.eps}}  % bb=108 0 508 585  %
%-->\fbox{\includegraphics[viewport=98 460 273 590,keepaspectratio,clip=true]{Rvdd-DE-9B2.eps}}  % bb=108 0 508 585  %
\fbox{\includegraphics[viewport=0 460 175 590,keepaspectratio,clip=true]{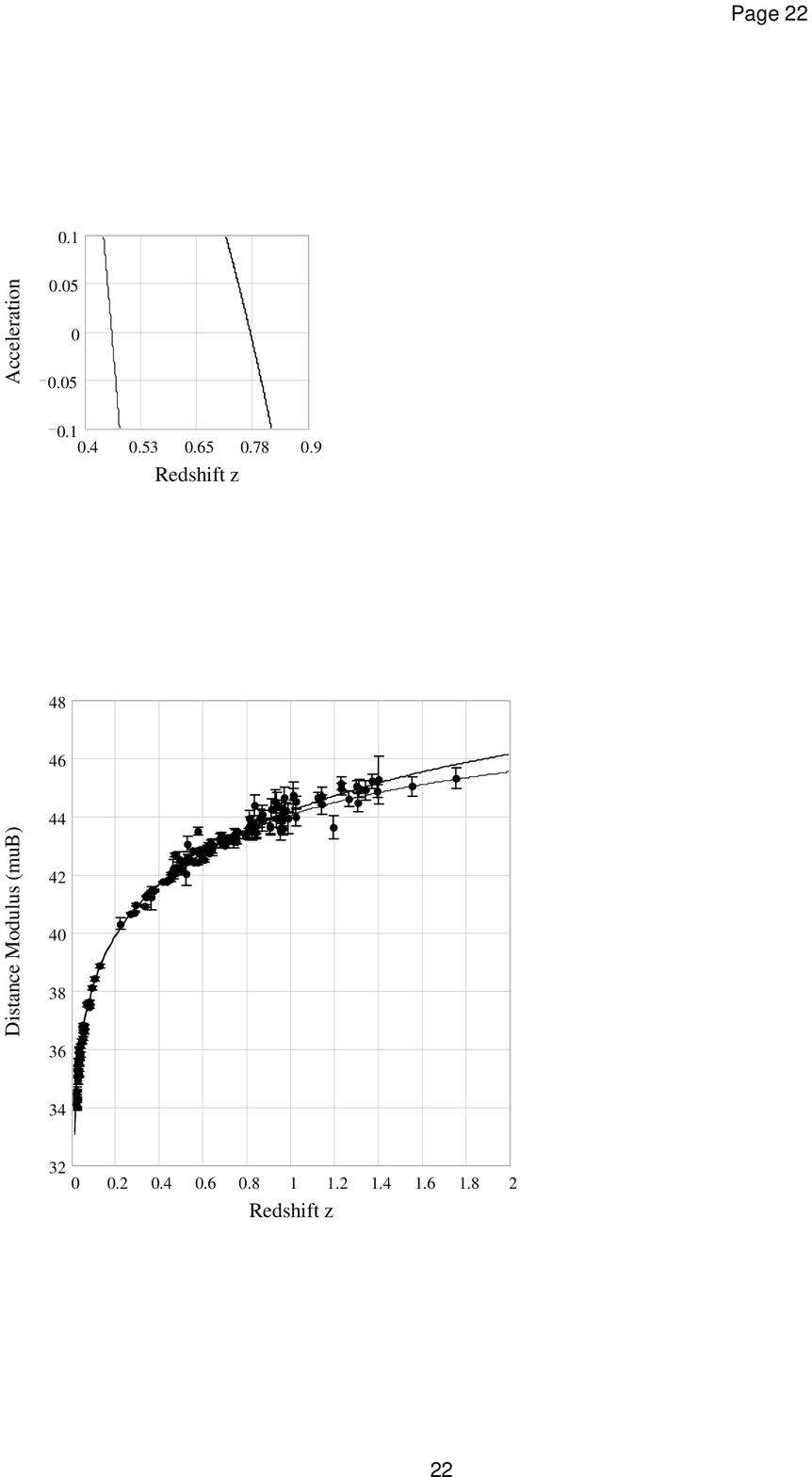}}  % bb=108 0 508 585  %
\caption{Scale factor acceleration $\ddot{R}_v$ at transition redshift.  Thin (left) curve is for
$\Omega_m = 0.3$, $\Omega_{\Lambda} = 0.7$,
$w_0 = -1.0$ and $w_a = 4.3$ with transition $z_t \approx 0.46$.   Thick (right) curve is for
$\Omega_m = 0.038$, $\Omega_{\Lambda} = 0.4$,
$w_0 = -1.0$ and $w_a = 4.3$ with transition $z_t \approx 0.78$.} 
\label{fig:scalfac-accel-trans-z-DarkMatterDarkEnergy}
\end{figure}

\begin{figure}[htb!]
%\centering %
\setlength{\fboxrule}{1.5pt}
%%%\fbox{\includegraphics[viewport=0 130 215 390,keepaspectratio,clip=true]{Rvdd-NoDE1.eps}}  % bb=108 0 508 585  %
%%%\fbox{\includegraphics[viewport=98 100 313 390,keepaspectratio,clip=true]{Rvdd-NoDE1-new.eps}}  % bb=108 0 508 585  %
%-->\fbox{\includegraphics[viewport=98 105 350 390,keepaspectratio,clip=true]{Rvdd-DE-9B2.eps}}  % bb=108 0 508 585  %
\fbox{\includegraphics[viewport=0 105 252 390,keepaspectratio,clip=true]{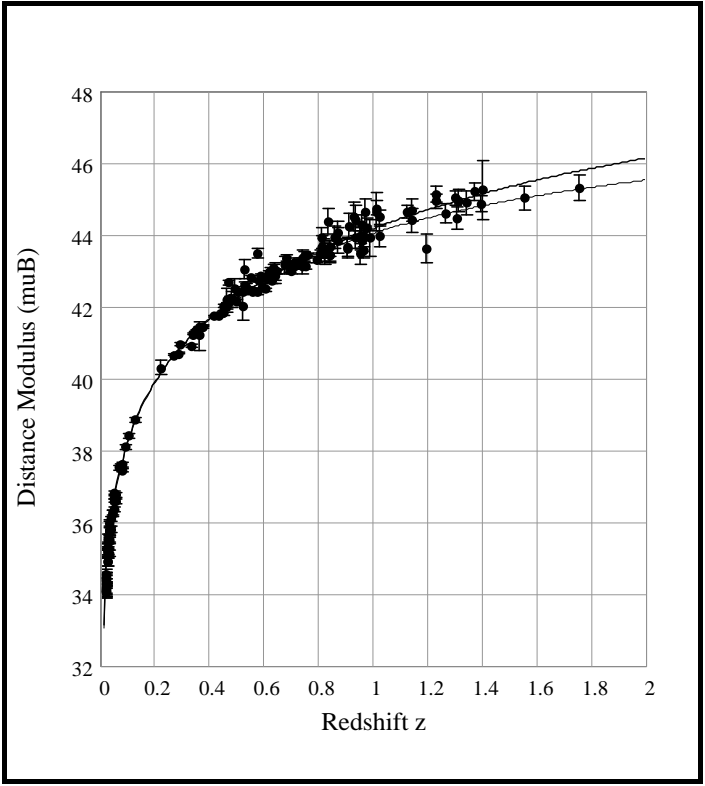}}  % bb=108 0 508 585  %
\caption{SNe Ia distance moduli. Points  are ($\mu_{\rm B} \pm \sigma_B$) from\cite{riess-2,astier}.
 The thick (upper) curve is for $\Omega_m = 0.038$, $\Omega_{\Lambda} = 0.4$, 
$w_0 = -1$, $w_a = 4.3$ and fitted offset $a = 0.064$ with $\chi^2/157 = 8.243$.
The thin (lower) curve is for $\Omega_m = 0.3$, $\Omega_{\Lambda} = 0.7$, 
$w_0 = -1$, $w_a = 4.3$ and  offset $a = -0.043$ with $\chi^2/157 = 8.626$.}
\label{fig:muB-vs-z-DarkMatterDarkEnergy}
\end{figure}

Fig. \ref{fig:scalfac-accel-trans-z-DarkMatterDarkEnergy} shows details of the acceleration $\ddot{R}_v$.
The thick curve is for $\Omega_m = 0.038$, $\Omega_{\Lambda} = 0.4$,
$w_0 = -1.0$ and $w_a = 4.3$ with transition $z_t \approx 0.78$. 
The thin curve is for $\Omega_m = 0.3$, $\Omega_{\Lambda} = 0.7$,
$w_0 = -1.0$ and $w_a = 4.3$ with transition $z_t \approx 0.46$ as was reported\cite{riess-1,riess-2}.
(We actually varied $w_a$ to get a smaller $\chi^2$ and in the process honed in on $z_t =0.46$.)
Fig. \ref{fig:muB-vs-z-DarkMatterDarkEnergy} shows the plots of the distance moduli for the SNe Ia data 
fitted to these two sets of parameters.  The thick (upper) curve is for $\Omega_m = 0.038$,
$\Omega_{\Lambda} = 0.4$,  $w_0 = -1$, $w_a = 4.3$ and fitted offset $a = 0.064$ with $\chi^2/157 = 8.243$.
The thin (lower) curve is for $\Omega_m = 0.3$, $\Omega_{\Lambda} = 0.7$, 
$w_0 = -1$, $w_a = 4.3$ and fitted offset $a = -0.043$ with $\chi^2/157 = 8.626$. 
The examples and parameters are summarized in Table \ref{tb:examples-1-2-3}.

For a brief summary, Example (Ex.) 2 has the smallest $\chi^2/N = 8.243$ where $N=157$, while Ex. 1 has the largest 
$\chi^2/N = 8.809$, with Ex. 3 in between with a $\chi^2/N = 8.626$.  However, both Ex.2 and Ex. 3 have exceptionally high
evolution ($w_0=-1.0$, $w_a=+4.3$) toward higher redshift, of what may be termed a ``anti-phantom'' energy, which might
be characterized as physically unlikely because it has not been observed so far. On the other hand, 
Ex. 1 has an evolution ($w_0=-1.0$, $w_a=+1.0$) of ``anti-dark'' energy toward higher redshift which cancels 
the dark energy development, which is physically more plausible since it involves dark energy for which there is observable
evidence. Thus, qualitatively, Ex. 1 is given a ``best fit" rating overall.
Ref. (Riess, et al.\cite{riess-1}, Sect. 4.3) for a nice exposition on dark energy.

These three examples provide a glimpse at the cosmology.  Consider that there are $N_C=2 \times 3^3 = 54$
possible categories of the parameter space:  
$\Omega_m$ with no dark matter or dark matter,
$\Omega_{\Lambda}$,  $w_0$  and $w_a$  each either negative, zero or positive.
Obviously, systematic analyses and larger data sets are required to narrow down the parameter space.

\begin{table}
\caption{Curve fit parameters.}
\begin{tabular}{| c | c | c | c | c | c | c| c | c |}
\hline
Ex. & $\Omega_m$ & $\Omega_{\Lambda}$ &  $K c \tau$ & $w_0$ & $w_a$ & $a$ & $\chi^2 / 157$ & $z_t$ \\
\hline
1 & 0.038 & -0.019 & -0.981 & -1.0 & +1.0 & +0.164 & 8.809 & 0.84 \\
\hline
2 & 0.038 & +0.4 & -0.562 & -1.0 & +4.3 & +0.064 & 8.243 & 0.78 \\
\hline
3 & 0.3 & +0.7 & 0 & -1.0 & +4.3 & -0.043 & 8.626 & 0.46 \\
\hline
\end{tabular}
\label{tb:examples-1-2-3}
\end{table}

\section{Carmeli Cosmology \label{sec:carmeli-cosmology} }

The 4-D space-velocity cosmology of Carmeli\cite{carmeli-0,behar-carmeli} is based on the premise  of a constant scale factor $R_v$.  Assume
\begin{eqnarray}
  \ddot{R}_v  &=&  \dot{R}_v = 0,  \label{eq:Rvdot:Rvddot=approx=1} \\
  \nonumber \\
  R_v  &=& 1,  \nonumber
\end{eqnarray}
with
\begin{eqnarray}
  R &=& R_r = r,   \label{eq:R-carmeli} \\
  \nonumber \\
  R' &=& R_r' =  1,  \label{eq:R-prime-carmeli}  \\
  \nonumber \\
  R'' &=& R_r'' =  0.  \label{eq:R-dub-prime-carmeli}
\end{eqnarray}
The key equations are (\ref{eq:basic-2b-sol-1b}), (\ref{eq:cal-F-equation})  and (\ref{eq:f-differential}).  
With the values Eqs. (\ref{eq:R-carmeli})-(\ref{eq:R-dub-prime-carmeli}) substituted into (\ref{eq:cal-F-equation}) 
we get
\begin{equation}
 -\kappa \tau^2 \rho_{eff}  = \frac{ f^{'} } { r }  + \frac{ f } { r^2 }  =  F_o,
    \label{eq:carmeli-F_o-equation}
\end{equation}
where $F_o$ is a constant  since $\rho_{eff}$ is assumed to be a constant with respect to $v$ and $r$.
However, it is assumed that $\rho_{eff}$ is a function of the cosmic time.
For the inhomogeneous case where $F_o \ne 0$ the solution of (\ref{eq:carmeli-F_o-equation}) is
\begin{equation}
  f \left( r \right) = f_o  r^2,  \label{eq:f-carmeli-inhomo-sol}
\end{equation}
where $f_o$ is a constant.
And, for the homogeneous case where $F_o = 0$ the solution is
\begin{equation}
  f \left( r \right) = - \frac{ 2 G M } {c ^2 r },  \label{eq:f-homo-carmeli-sol}
\end{equation}
where $M$ is a constant point mass centered at the origin of coordinates.
The general solution is the sum of (\ref{eq:f-carmeli-inhomo-sol}) and (\ref{eq:f-homo-carmeli-sol}),
\begin{equation}
   f \left( r \right) = f_o  r^2  - \frac{ 2 G M } {c ^2 r },  \label{eq:f-general-carmeli-sol}
\end{equation}
where the system of coordinates is centered on the central mass $M$.
Using (\ref{eq:f-general-carmeli-sol}) in (\ref{eq:carmeli-F_o-equation})  gives the value
\begin{equation}
  F_o = 3 f_o.  \label{eq:carmeli-Fo-val}
\end{equation}
From (\ref{eq:basic-2b-sol-1b}), with $\ddot{R}_v = \dot{R}_v = 0$ and $R_v = 1$ we get
\begin{equation}
   \kappa \frac{\tau}{c} p_{eff}  =  \frac{f\left( r \right) }{r^2}  =  f_o. 
            \label{eq:carmeli-basic-1}
\end{equation}
Substituting for $f(r)$ from (\ref{eq:f-general-carmeli-sol}) into (\ref{eq:carmeli-basic-1}) gives
the effective pressure
\begin{equation}
  p_{eff}  =  \frac{ c } { \kappa \tau } f_o - \frac{ c \tau M } { 4 \pi r^3 }.   \label{eq:carmeli-pressure-val}
\end{equation}
The value of $f_o$ is obtained from (\ref{eq:basic-1b-sol-1b})  with $\dot{R}_v = 0$ and $R_v = 1$,
so that at $v=0$ it gives us
\begin{equation}
  f_o =  - \frac{1}{3} \kappa \tau^2 \rho_{eff}.   \label{eq:f_o-rho_eff-carmeli}
\end{equation}
With this value for $f_o$ we obtain from (\ref{eq:carmeli-pressure-val}) the effective pressure
\begin{equation}
   p_{eff}  =  - \frac{1} {3} c \tau \left( \rho_{eff}  + \rho_M\left( r \right) \right), 
       \label{eq:basic-2b-sol-carmeli}
\end{equation}
where
\begin{equation}
  \rho_M( r )  =  \frac{ 3 M } { 4 \pi r^3 }  \label{eq:carmeli-central-mass-density}
\end{equation}
is the central mass density, which at this point is a convenient mathematical construct.
The function (\ref{eq:int-mu-1c}) for $e^{\mu}$ is now given by  
\begin{equation}
  e^{\mu} =  \frac{ 1 }{ 1 + f\left( r \right) } = \frac{ 1 }
         { 1 -  \frac{1}{3} \kappa \tau^2  r^2 \rho_{eff}  - 2 G M / c ^2 r  },
        \label{eq:int-mu-carmeli}
\end{equation}
where
\begin{equation}
   1 -  \frac{1}{3} \kappa \tau^2  r^2 \rho_{eff}  - 2 G M / c ^2 r   > 0.  \label{eq:carmeli-f(r)-boundary-condtion}
\end{equation}
We make note that Carmeli did not define an effective pressure, but simply used an ordinary pressure.
In our analysis the effective pressure is used.  We could setup an equation of state for Carmeli cosmology that
is defined, taking the central mass $M=0$ in (\ref{eq:basic-2b-sol-carmeli}),
\begin{equation}
  p_{eff}  =  w  c \tau \rho_{eff},  \label{eq:carmeli-equation-of-state}
\end{equation}
where $w = -1/3$.  Looking at (\ref{eq:p-w-rho})  this implies that $w_0 = -1/3$ and  $w_a=0$.

In Carmeli cosmology the effective mass density $\rho_{eff} = \rho - \rho_c$, so we equate it with our definition
 (\ref{eq:rho_eff_fin}) and obtain
\begin{equation}
  \rho_{eff}  =  \rho_m  + \rho_{\Lambda}  =  \rho - \rho_c  ,   \label{eq:carmeli-rho_eff}
\end{equation}
from which, with $\rho_m = \rho$,  we get
\begin{equation}
  \rho_{\Lambda}  =   - \rho_c,    \label{eq:carmeli-rho_lambda}
\end{equation}
where $\rho_c = 3 / 8 \pi G \tau^2$ is the critical density defined by (\ref{eq:rho_c_def}).
Using (\ref{eq:rho_V-def}) this yields
\begin{eqnarray}
    \rho_{\Lambda}  &=&  - \frac{  \Lambda } { \kappa \tau^2 }  =  - \rho_c,   \label{eq:rho_lambda=-rho_c}  \\
  \nonumber \\
   \Lambda   &=&  \kappa \tau^2  \rho_c  =   \frac{ 3 } { c^2 \tau^2 }.    \label{eq:lambda-value}
\end{eqnarray}
Allowing for the $c^2$ this is the relation for $\Lambda$ which was reported\cite{carmeli-kuzmenko}.
However, this does not fix the value of $\Lambda$ but it appears to experimentally have the same order of magnitude\cite{behar-carmeli}.  Using (\ref{eq:carmeli-rho_eff}) and (\ref{eq:carmeli-rho_lambda}), the effective mass density parameter $\Omega_{eff}$ is given by
\begin{equation}
   \Omega_{eff}  = \frac{\rho_m + \rho_{\Lambda}} {\rho_c} 
                         = \Omega_m + \Omega_{\Lambda} ,        \label{eq:Omega_eff_carmeli}
\end{equation}
where 
\begin{equation}
   \Omega_m = \frac{ \rho_m } { \rho_c }   \label{eq:omega_m-carmeli}
\end{equation}
and
\begin{equation}
   \Omega_{\Lambda}  = \frac{ \rho_{\Lambda} } { \rho_c }   = \frac{ -\rho_c} { \rho_c }  = -1. \label{eq:omega_lambda-carmeli}
\end{equation}
With no central mass, $M=0$, and using Eqs. (\ref{eq:Omega_eff_carmeli})-(\ref{eq:omega_lambda-carmeli})
then (\ref{eq:int-mu-carmeli}) takes the form
\begin{equation}
   e^{\mu} =  \frac{ 1 }{ 1   -  \left(  \Omega_{eff} \right) r^2 / c^2 \tau^2 } 
                 =  \frac{ 1 }{ 1   +  \left(  1 - \Omega_m  \right) r^2 / c^2 \tau^2 }.  \label{eq:int-mu-carmeli-form-1}
\end{equation}
Equation (\ref{eq:int-mu-carmeli-form-1}) is written in that form because $\Omega_m < 1$ at the present epoch.
The curvature $K = (\Omega_m  - 1 ) / c^2 \tau^2 < 0$ which implies that the Carmeli universe has a 
hyperbolic (open) spatial geometry.   

Setting $ds = 0$ in (\ref{eq:dr/dv-0}) for the expansion of the universe gives the differential equation
\begin{equation}
  \tau dv  =  \frac{ dr } {  \sqrt{ 1   +  \left(  1 - \Omega_m \right) r^2 / c^2 \tau^2  }  }.  \label{eq:expansion-carmeli}
\end{equation}
Upon integration, assuming $1 - \Omega_m  > 0$, (\ref{eq:expansion-carmeli}) yields for the expansion velocity
\begin{equation}
   v  =   \frac{ c \, {\rm sinh}^{-1} \left(  \sqrt{ 1 - \Omega_m } r / c \tau \right) } {  \sqrt{ 1 - \Omega_m }  }.
            \label{eq:expan-carmeli-solve}
\end{equation}
Inverting (\ref{eq:expan-carmeli-solve}) we obtain the velocity-distance relation
\begin{equation}
   r = \frac{ c \tau \, {\rm sinh} \left(  \sqrt{ 1 - \Omega_m } v / c   \right)  } {  \sqrt{ 1 - \Omega_m }  }.  \label{eq:r-vs-v-carmeli}
\end{equation}

When the central mass $M > 0$, the expression (\ref{eq:int-mu-carmeli}) takes the form
\begin{equation}
   e^{\mu} =  \frac{ 1 }{ 1 + \left(  1 - \Omega_m \right) r^2 / c^2 \tau^2  -  2 G M / c^2 r  }.
         \label{eq:int-mu-carmeli-form-2}
\end{equation}
Then the differential equation for the universe expansion (\ref{eq:dr/dv-0}) takes the form
\begin{equation}
  \tau dv  =  \frac{ dr } {  \sqrt{ 1 + \left( 1 - \Omega_m \right) r^2 / c^2 \tau^2  - 2 G M / c^2 r  }  }. 
         \label{eq:expansion-carmeli-point-mass}
\end{equation}
The velocity-distance relation is then given by
\begin{equation}
    v = h \int^r_0 { \frac{ dr' } {  \sqrt{ 1 + \left( 1 - \Omega_m \right) r'^2 / c^2 \tau^2  - 2 G M / c^2 r'  }  } },
       \label{eq:v-distance-integral}
\end{equation}
where the expansion is centered on the point mass $M$.  This model was recently described by Hartnett\cite{hartnett-5}.

This is the basics of the cosmology.  For $\Omega_m < 1$  the curvature $K = (\Omega_m - 1 ) / c^2 \tau^2 < 0$ defines an open universe. 
Since  the scale factor $R_v$ is assumed to be constant,  a velocity-redshift
 relation must be obtained by other methods.  The special relativistic Doppler velocity-redshift relation
\begin{equation}
  \frac{v}{c} = \frac{ \left( 1 + z \right)^2 - 1 } { \left( 1 + z \right)^2 + 1 }  \label{eq:carmeli-Doppler-vel-z}
\end{equation}
 is often used.  However, since here the scale factor $R_v$ is constant, the redshift $z$ is assumed
to be a function of the cosmic time.
Also, the evolution of the mass density parameter $\Omega_m$ is assumed to be in the time domain
and its variation also must be obtained by other methods.  A further limitation is that in (\ref{eq:expansion-carmeli}),
 the restriction that $1 + \left(  1 - \Omega_m \right) r^2 / c^2 \tau^2 > 0$  requires $\Omega_m  \le  2$
approximately.  This makes  it difficult for high redshift data analysis\cite{hartnett-oliveira,hartnett-1}.
On the other hand, when cosmic time is added as a fifth dimension the CGR 5-D time-space-velocity model is applicable
to galaxy dynamics because the cosmological redshift across a galaxy region is nearly 
constant\cite{hartnett-2,carmeli-4,hartnett-3,hartnett-4}.

\section{Conclusion} 

A general solution to the Einstein field equations has been obtained for the  four dimensional space-velocity
Cosmological General Relativity theory of Carmeli.  This development provides the tools necessary for
the analysis of astrophysical data.  In particular, a redshift-distance relation is given, 
analogous to the standard FLRW cosmology, and an evolving equation of state is provided.
An analysis of high redshift SNe Ia data was made to show the efficacy of the cosmology. Model examples
were given with only baryonic matter and dark energy in hyperbolic space and with baryonic plus dark matter and 
dark energy in flat space. Finally, Carmeli's cosmology in 4-D was obtained.

\acknowledgement

   The author is grateful  to the anonymous reviewers for their suggestions.


\begin{thebibliography}{}

   \bibitem{carmeli-0} Carmeli, M.:
      Relativity: Modern Large-Scale Spacetime Structure of the Cosmos.
      World Scientific, Singapore (2008)

   \bibitem{behar-carmeli} Behar, S., and Carmeli, M.:
      Cosmological relativity: a new theory of cosmology.
      Int. J. Theor. Phys.  {\bf 39}(5), 1375-1396 (2000). arXiv:astro-ph/0008352

   \bibitem{carmeli-1} Carmeli, M.:
      Cosmological general relativity.
      Commun. Theor. Phys. {\bf 5}, 159 (1996)

     \bibitem{hartnett-2} Hartnett, J. G.: 
      The Carmeli metric correctly describes spiral galaxy rotation curves.
      Int. J. Theor. Phys. {\bf 44}, 359 (2005). arXiv:gr-qc/0407082

   \bibitem{oliveira-1} Oliveira, F. J.:
      Particle pair production in cosmological general relativity.
      Int. J. Theor. Phys. {\bf 51}(12), 3993-4005 (2012). arXiv:1203.4797
 
   \bibitem{bekenstein-1} Bekenstein, J. D.:
      Black holes and entropy.
      Phys. Rev. D {\bf 7}(8), 2333-2346 (1973). doi:10.1103/PhysRevD.7.2333

   \bibitem{hawking-1} Hawking, S. W.:
      Particle creation by black holes.
      Commun. Math. Phys. {\bf 43}(3), 199-220 (1975). doi:10.1007/BF02345020

   \bibitem{fang} Fang, W., Hu, W., and Lewis, A.:
      Crossing the phantom divide with parameterized post-friedmann dark energy. 
      Phys. Rev. D {\bf 78}, 087303 (2008). arXiv:0808.3125

   \bibitem{tolman-1}Tolman, R. C.:
      Relativity Thermodynamics and Cosmology.
      Dover, New York (1987)
    
   \bibitem{oliveira-hartnett-0}Oliveira, F. J., and Hartnett, J. G.:
      Carmeli's cosmology fits data for an accelerating and decelerating universe without dark matter or dark energy.
      Found. Phys. Lett. {\bf 19}(6), 519-535 (2006). arXiv:astro-ph/0603500 

   \bibitem{riess-1}  Riess, A. G., et al.:
      Type Ia supernova discoveries at $z > 1$ from the hubble space telescope: evidence for
      past deceleration and constraints on dark energy evolution.
      Astrophs. J. {\bf 607}, 665-687 (2004)

   \bibitem{riess-2} Riess, A. G., et al.:
      New hubble space telescope discoveries of type Ia supernovae at $z \ge 1$: 
      narrowing constraints on the early behavior of dark energy.
      Astrophs. J. {\bf 659}(1) (2007).  doi:10.1086/510378.
      arXiv:astro-ph/0611572(2006)

   \bibitem{astier} Astier, P., et al.:
      The supernova legacy survey: measurement of  $\Omega_{\mathsf{M}}$,
      $\Omega_\mathsf{\Lambda}$ and $w$ from the first year data set.
      Astro. \& Astrophs. {\bf 447}(1), 31-48 (2006).
      doi: 10.1051/0004-6361:20054185.  arXiv:astro-ph/0510447

   \bibitem{burles-1} Burles, S., Nollett, K. M., and Turner, M. S.:
     Big bang nucleosynthesis predictions for precision cosmology.
    Astrophs. J. {\bf 552}, L1-L5 (2001)
 
   \bibitem{carmeli-kuzmenko}  Carmeli, M., and Kuzmenko, T.:
      Value of the cosmological constant in the cosmological relativity theory.
      Int. J. Theor. Phys. {\bf 41}(1), 131-135 (2002). arXiv:astro-ph/0110590

   \bibitem{hartnett-oliveira} Hartnett, J. G., and Oliveira, F. J.:
      Luminosity distance, angular size and surface brightness in cosmological general relativity.
      Found. Phys.  {\bf 37}(3), 446-454 (2007a). arXiv:astro-ph/0603500

   \bibitem{hartnett-1} Hartnett, J. G.:
      Extending the redshift-distance relation in cosmological general relativity to higher redshifts.
      Found. Phys.  {\bf 38}(3), 201-215 (2008). doi:10.1007/s10701-007-9198-5

    \bibitem{hartnett-5} Hartnett, J.G.:
      A valid finite bounded expanding Carmelian universe without dark matter.
      Int. J. Theor. Phys. {\bf 52}(12): 4360-4366 (2013)

    \bibitem{carmeli-4}  Carmeli, M.:
      Is galaxy dark matter a property of spacetime?
      Int. J. Theor. Phys. {\bf 37}(10), 2621-2625 (1998)

      \bibitem{hartnett-3} Hartnett, J.G.:
        Spiral galaxy rotation curves determined from Carmelian general relativity.
       Int. J. Theor. Phys. {\bf 45}(11), 2147-2165 (2006)

      \bibitem{hartnett-4} Hartnett, J.G.:
        Spheroidal and elliptical galaxy radial velocity dispersion determined from cosmological general relativity.
        Int. J. Theor. Phys. {\bf 47}(5), 1252-1260 (2008)

\end{thebibliography}
\end{document}